\documentclass[journal]{IEEEtran}
\usepackage[T1]{fontenc}
\usepackage[utf8]{inputenc}
\usepackage{color}
\usepackage{array}
\usepackage{float}
\usepackage{multirow}
\usepackage{amsmath}
\usepackage{graphicx}
\usepackage[bookmarks=true,bookmarksnumbered=true,bookmarksopen=true,bookmarksopenlevel=1,
 breaklinks=false,pdfborder={0 0 0},pdfborderstyle={},backref=false,colorlinks=false]
 {hyperref}
\hypersetup{pdftitle={Your Title},
 pdfauthor={Your Name},
 pdfpagelayout=OneColumn, pdfnewwindow=true, pdfstartview=XYZ, plainpages=false}

\makeatletter

\providecommand{\tabularnewline}{\\}
\newcommand{\lyxdot}{.}

\floatstyle{ruled}
\newfloat{algorithm}{tbp}{loa}
\providecommand{\algorithmname}{Algorithm}
\floatname{algorithm}{\protect\algorithmname}

\usepackage{amsmath,amsfonts}
\usepackage{cite}

\usepackage{array}
\usepackage{textcomp}
\usepackage{stfloats}
\usepackage{gensymb}
\usepackage{url}
\usepackage{verbatim}
\usepackage{graphicx}
\usepackage{balance}
\usepackage{babel}

\usepackage[labelformat=simple,caption=false,font=footnotesize]{subfig} 

\usepackage{diagbox}

\usepackage{algorithm}
\usepackage{algorithmicx}
\usepackage{algpseudocode}

\ifdefined\showcaptionsetup
 \PassOptionsToPackage{caption=false}{subfig}
\fi
\usepackage{subfig}
\makeatother

\begin{document}
\title{BagChain: A Dual-functional Blockchain Leveraging Bagging-based Distributed
Learning}
\author{Zixiang Cui, Xintong Ling,~\IEEEmembership{Member,~IEEE}, Xingyu
Zhou, Jiaheng Wang,~\IEEEmembership{Senior Member,~IEEE}, Zhi~Ding,~\IEEEmembership{Fellow,~IEEE},
Xiqi Gao,~\IEEEmembership{Fellow,~IEEE}
\thanks{Zixiang Cui, Xintong Ling, Xingyu Zhou, Jiaheng Wang, and Xiqi Gao are with the National Mobile Communications Research Laboratory, Southeast University, Nanjing 210096, China (email: zxcui@seu.edu.cn; xtling@seu.edu.cn; zhouxingyu319@gmail.com, jhwang@seu.edu.cn; xqgao@seu.edu.cn). Xintong Ling, Jiaheng Wang, and Xiqi Gao are also with the Purple Mountain Laboratories, Nanjing 210023, China.}
\thanks{Zhi Ding is with the Department of Electrical and Computer Engineering, University of California at Davis, Davis, CA 95616 USA (e-mail: zding@ucdavis.edu).}}
\maketitle
\begin{abstract}
This work proposes a dual-functional blockchain framework named BagChain
for bagging-based decentralized learning. BagChain integrates blockchain
with distributed machine learning by replacing the computationally
costly hash operations in proof-of-work with machine-learning model
training. BagChain utilizes individual miners' private data samples
and limited computing resources to train base models, which may be
very weak, and further aggregates them into strong ensemble models.
Specifically, we design a three-layer blockchain structure associated
with the corresponding generation and validation mechanisms to enable
distributed machine learning among uncoordinated miners in a permissionless
and open setting. To reduce computational waste due to blockchain
forking, we further propose the cross fork sharing mechanism for practical
networks with lengthy delays. Extensive experiments illustrate the
superiority and efficacy of BagChain when handling various machine
learning tasks on both independently and identically distributed (IID)
and non-IID datasets. BagChain remains robust and effective even when
facing resource-constrained mobile devices, heterogeneous private
user data, and sparse wireless network connectivity.
\end{abstract}

\begin{IEEEkeywords}
Bagging, blockchain, consensus protocol, distributed learning, proof-of-useful-work.\vspace{-0.1cm}
\end{IEEEkeywords}

\section{Introduction}

The proliferation of mobile devices that collect data with powerful
sensors like cameras and microphones enables the possibility of constructing
more powerful intelligent applications. However, the centralized machine
learning (ML) approach is facing certain challenges in exploiting
private data and computing power on mobile devices. First of all,
data transmissions from users to data centers over bandwidth-limited
networks are time-consuming and costly. Second, users are also unwilling
to share their sensitive information with data centers, precluding
them from harnessing such data for ML purposes. Third, users' private
data may be non-independently and identically distributed (non-IID),
which significantly increases the bias of the ML models trained with
mobile devices' private data \cite{He2009}. Last but not least, mobile
devices lack the computing capability required to train high-quality
ML models.

Some distributed learning techniques like federated learning (FL)
\cite{McMahan2017} and consensus learning (CL) \cite{Georgopoulos2014}
can harness computing resources and heterogeneous private data on
mobile devices to train ML models. In FL systems, data holders train
local models with their private data, and the local models are then
uploaded and aggregated into a global model by a central server. CL
is a fully decentralized approach where each node updates the local
model by aggregating the model parameters received from the neighboring
nodes in a mesh network. 

However, cooperative ML for on-device distributed learning faces several
challenges when migrating existing distributed learning approaches
to peer-to-peer open networks, which may include untrustworthy or
faulty nodes. In some distributed learning techniques like FL, centralized
network topology incurs single-point-of-failure vulnerability and
communication bottlenecks at the central server \cite{Kairouz2021}.
The iterative training procedures in FL also incur huge communication
overhead, which is unbearable for both mobile devices and central
servers. Other techniques like CL assume a fixed network topology,
which is not feasible in dynamic network environments allowing nodes
to join or leave frequently. Currently, most distributed learning
techniques are set up in trustworthy closed systems where data access
and contributions from external nodes are prohibited, while the open
and untrustworthy environment may pose security risks like poisoning
attacks to these methods. 

Blockchain is a game-changing innovation that subverts the conventional
centralized paradigm and has shown its potential for various applications
\cite{Ling2020,Cao2024,Ling2025}. It exploits consensus protocols
and cryptographic tools to store data consistently and immutably in
an open and untrustworthy network. Compared with centralized solutions,
blockchain is more scalable and resilient to single-point failures
and malicious entities.

Serving as a decentralized foundation, blockchain is capable of offering
promising solutions to the challenges in existing distributed learning
techniques.  Each node executes consensus protocols and validates
on-chain data independently to mitigate the risk of malfunctioning
or dishonest nodes compared with centralized ML systems. With the
help of blockchain, ML systems are more friendly to external participants
than conventional ML systems because  participants can join the distributed
learning process freely without being permitted or managed by any
central node.  Moreover,  blockchain  can incentivize participants
to   contribute their computing power and private data for distributed
ML model training.\vspace{-0.1cm}

\subsection{Related Work\label{sec:Related-Work}\vspace{-0.1cm}}

Since 2019, numerous studies have adapted blockchain systems to various
ML tasks such as biomedical image segmentation model training \cite{li2020dlbc},
cloud–edge–end resource allocation \cite{Qiu2021}, neural architecture
searching \cite{LI2022100089}, and medical image fusion \cite{Xiang2023}
to leverage computing power and data in a blockchain network.  
Several studies \cite{BravoMarquez2019,Chenli2019,Kang2020,Li2021byzantine,Chai2021,Liu2021}
utilized blockchain as an ML competition platform, where ML model
training, evaluation, and ranking procedures are embedded in the workflow
of the blockchain system. In such systems, model checkers filter
out unqualified model parameters or updates based on performance evaluation
on a test dataset provided by the model requester, and the node that
trains the best ML model within the least time sends its model to
the model requester and receives the training reward. However, the
above approaches, which select and reward only one single winning
model from several models trained on specific datasets and discard
the rest, fail to take full advantage of miners' computing power and
private data samples.

As a collaborative ML approach that can harness device-side private
data in a privacy-preserving manner, FL can be blended with blockchain
systems better than the aforementioned blockchain-based ML competition
schemes. Several studies implemented FL algorithms in blockchain systems
via smart contracts \cite{Toyoda2019,Ramanan2020}. A more fundamental
approach is integrating FL with customized consensus protocols so
that the FL system can become self-incentivized and more robust to
malicious behaviors. For instance,  in proof-of-federated-learning
(PoFL) \cite{Qu2021}, Qu \emph{et al.} adapted the mining pools in
proof-of-work (PoW) blockchain into federations of miners, where the
miners in a mining pool collaboratively train a global model with
algorithms like federated averaging and the pool manager broadcasts
the block containing the global model.  In \cite{Wang2022}, the
authors eliminated the trusted third-party platform in PoFL with a
new block structure, extra transaction types, and a credit-based incentive
mechanism. In FedCoin \cite{Liu2020fedcoin}, a Shapley-value-based
consensus protocol, the quality of local models was considered before
allocating training rewards by evaluating the contribution of each
miner's local model to the global model.  In \cite{Cao2023,Ying2023},
the authors combined directed acyclic graph (DAG) blockchain with
FL to avoid the drain of computing resources in PoW systems and protect
FL from lazy nodes and poisoning attacks.   However, in many blockchain-based
FL systems, the FL iterations are administered by a few centralized
servers (e.g., the mining pool managers in PoFL) or  third-party
platforms, which compromises the degree of decentralization of the
entire system. Though this issue has been considered in the above
two works using DAG-based asynchronous FL, the low utilization of
local models affects the convergence rate of the global model and
degrades the learning performance \cite{Zhang2024c}.

Highlight that ML can serve as useful work in proof-of-useful-work
(PoUW) as an alternative to PoW and improves the environmental sustainability
of conventional PoW systems by replacing the computing power consumed
in meaningless hash calculations with some useful computation tasks
like ML model training. Most ML-based PoUW protocols like proof-of-deep-learning
(PoDL) \cite{Chenli2019}, personalized artificial intelligence \cite{Zhang2023jul},
proof-of-learning (PoL) \cite{Jia2021}, RPoL \cite{Zhang2023jul},
and distributed PoDL \cite{Su2023} use the same principle where miners
attest certain amounts of honest computation by providing the intermediate
parameters necessary for reproducing the final ML model. Such protocols
involve intense communication and computation overhead because  ML
model retraining or verifying intermediate training steps are necessary
when validating miners' workloads. To avoid transmitting all the intermediate
model parameters directly, in DLchain \cite{Chenli2020}, miners execute
training algorithms like stochastic gradient descent with fixed random
seeds and prove training work with the random seeds and the accuracies
of the model at each epoch.  Also, zero-knowledge PoL \cite{Zhang2024a}
significantly reduced the communication, computation, and storage
overhead for Internet of Things devices via zero-knowledge proof techniques.
Nevertheless, even  so, the computational amount involved in PoUW
proof generation or validation remains unacceptable  for resource-constrained
mobile devices.\vspace{-0.1cm}

\subsection{Our Contributions\vspace{-0.1cm}}

In our work, we propose BagChain by integrating the blockchain protocol
with bagging, an ensemble learning algorithm \cite{Breiman_1996}.
BagChain serves as a dual-functional blockchain framework targeting
two purposes together: distributed machine learning and blockchain
consensus maintenance. Specifically, we design the three-layer blockchain
structure for BagChain and utilize the computational power for useful
ML model training instead of meaningless hash operations in PoW. It
features a fully decentralized architecture without relying on central
nodes or trusted third parties. In BagChain, all miners train the
possibly weak base models by using their local computing power and
private data and aggregate the base models into an ensemble model,
which exhibits satisfactory performance according to the experimental
results in Section \ref{sec:ChainXim-based-Simulation}. Notably,
these base models are trained on miners' private datasets, obviating
the need to reveal or share raw data, and BagChain can tolerate these
private datasets being imbalanced or non-IID. Furthermore, each base
model could be very weak since it requires only a short training period
of 10 to 20 epochs, and thus BagChain is feasible even for resource-constrained
mobile nodes. Furthermore, since the bagging algorithm only requires
one-time prediction-level aggregation rather than the iterative global
model aggregation in FL, BagChain has a lower communication overhead
than vanilla FL and is thus suitable for bandwidth-limited wireless
networks. We have released the code of BagChain at: https://github.com/czxdev/BagChain.

The main contributions of our work are summarized below:
\begin{itemize}
\item We propose BagChain, a dual-functional blockchain framework that enables
generic ML task execution and integration of the bagging algorithm
in a public blockchain network. To the best of our knowledge, it is
the first work integrating blockchain into bagging-based distributed
learning algorithms.
\item We propose the three-layer blockchain structure of BagChain with block
generation and validation algorithms to enable the automation of model
training, aggregation, and evaluation in a fully decentralized setting.
\item We introduce a cross fork sharing mechanism to reduce computing power
waste resulting from blockchain forking. The mechanism maximizes the
utilization of the base models and improves the performance of the
ensemble models by including the hash values of as many MiniBlocks
in Ensemble Blocks as possible.
\item We discuss the performance issues and security implications of BagChain
including model plagiarism and blockchain forking, and design a task
queuing mechanism to improve BagChain's resilience to blockchain forking.
\item We conduct ChainXim-based simulations for BagChain on the commonly
used image classification tasks to demonstrate the benefits of exploiting
non-IID private data and pooling computing power simultaneously and
to validate the feasibility of BagChain under different network conditions.
\end{itemize}

The remaining part of the paper is organized as follows.  In Section
\ref{sec:System-Model}, we present the system model. We show the
framework of BagChain and the validation procedures in Section \ref{sec:System-Design}
and Section \ref{sec:Block-Validation}, respectively. We design the
cross fork sharing mechanism in Section \ref{sec:Cross-Fork-Sharing}.
In Section \ref{sec:Security-and-Perforance}, we discuss the security
and performance. In Section \ref{sec:ChainXim-based-Simulation},
we provide the results of ChainXim-based experiments to demonstrate
the efficacy of BagChain. Section \ref{sec:Conclusion} draws a conclusion.\vspace{-0.1cm}

\section{System Model\label{sec:System-Model}}

\subsection{Distributed Learning Scenario\label{subsec:Distributed-Learning-Scenario}\vspace{-0.1cm}}

In this work, we consider a distributed learning scenario where some
requester nodes have demands for ML-based intelligent applications
and can offer financial incentives in exchange for ML model training
services. Meanwhile, some nodes, referred to as miners, are willing
to contribute their private data and computing power to ML tasks published
by these requesters. However, miners are unwilling to share their
private data due to privacy concerns and might have difficulty training
high-quality ML models independently since they may be mobile devices
with limited capabilities. Besides executing ML tasks, these miners
also need to maintain the blockchain network by generating new blocks.

In the considered scenario, we assume the label or feature distributions
across different miners' private datasets can be heterogeneous. Meanwhile,
we assume all requesters follow the rules of our framework, e.g.,
publishing sound tasks and releasing datasets in time, because they
have paid training fees when publishing ML tasks and lack the motivation
to disrupt the ML training process.\vspace{-0.1cm}

\subsection{Supervised Learning Model\vspace{-0.1cm}}

In our work, every ML task employs a supervised learning approach
where ML models can be described by parameterized functions $f(\cdot;\boldsymbol{\omega})$
trained on a dataset $D$. Each ML model can make predictions with
the input features and is uniquely determined by the model framework
$f$ and the model parameters $\boldsymbol{\omega}$. Each dataset
$D$ is composed of a feature matrix and a label vector $(\boldsymbol{X},\boldsymbol{y})$.
Each row of $\boldsymbol{X}$ denotes the feature vector of a sample,
and the corresponding $\boldsymbol{y}$ represents the categorical
label of the sample for classification tasks. Four datasets are involved:
public training dataset $D_{T}=(\boldsymbol{X}_{T},\boldsymbol{y}_{T})$,
private dataset $D_{M_{i}}=(\boldsymbol{X}_{M_{i}},\boldsymbol{y}_{M_{i}})$
that miner $M_{i}$ possesses, validation dataset $D_{V}=(\boldsymbol{X}_{V},\boldsymbol{y}_{V})$,
and test dataset $D_{E}=(\boldsymbol{X}_{E},\boldsymbol{y}_{E})$.
The public training dataset $D_{T}$ is available to all the miners
and is published by the requester before ML task execution. Meanwhile,
every miner's private dataset may be non-IID and heterogeneous. Mathematically,
$\lvert D\rvert$ means the number of samples in the dataset $D$,
and $D_{1}\bigcup D_{2}$ represents the union of $D_{1}$ and $D_{2}$.
We focus on the learning process and omit data standardization and
the other pre-processing steps.
\begin{figure*}[t]
\begin{centering}
\includegraphics[scale=0.7]{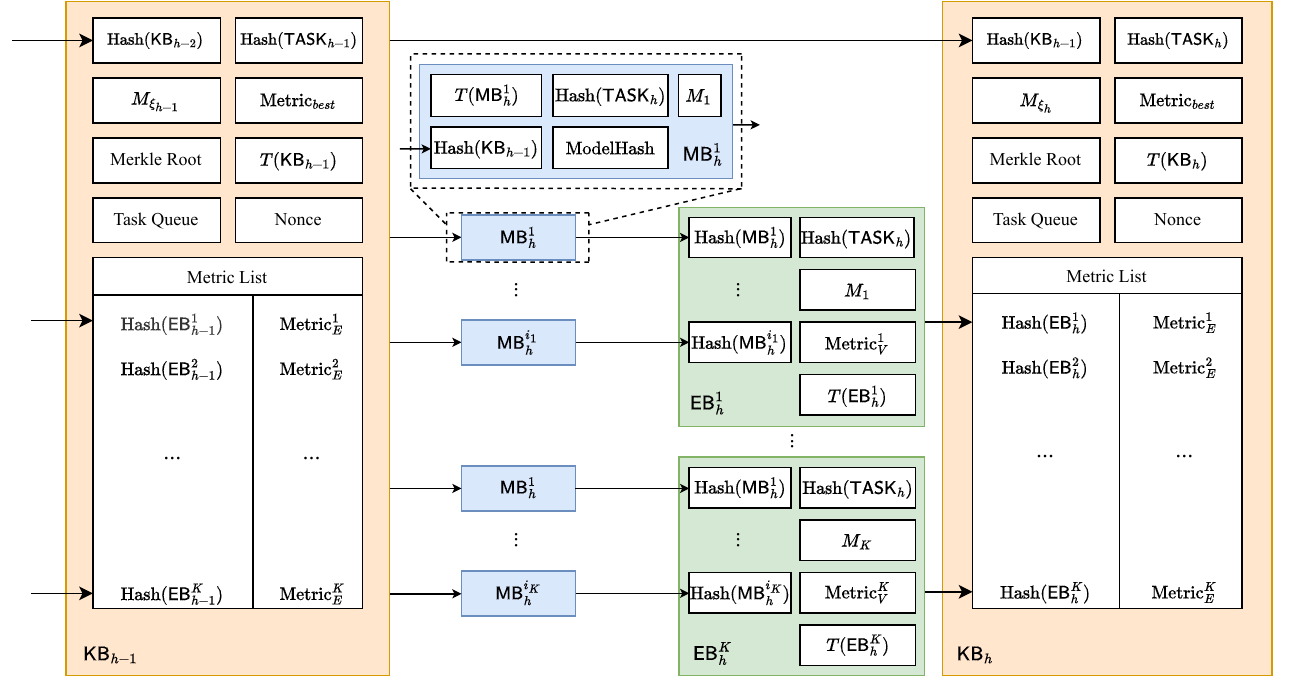}\vspace{-0.2cm}
\par\end{centering}
\begin{centering}
\caption{Overview of the proposed three-layer blockchain structure.\label{fig:Overview-of-data-structure}}
\par\end{centering}
\centering{}\vspace{-0.3cm}
\end{figure*}

The bagging algorithm involves four operations. The operation $\mathrm{Train}(\cdot;f)$,
parameterized by the model framework $f$, takes any dataset $D$
as input and outputs the model parameters $\boldsymbol{\omega}$ trained
on $D$. The operation $\mathrm{Resample(\cdot)}$ performs random
sampling with replacement on the input dataset. The operation $\mathrm{Aggregate(\cdot)}$
takes the output values of a sequence of base models as input and
outputs the final predictions of the ensemble of the base models using
a specific aggregation rule, typically a majority vote for classification
tasks. The operation $\mathrm{Metric}(\cdot,\cdot)$ takes the predicted
labels and the correct labels as inputs and outputs the performance
metric that evaluates the discrepancies between the two input vectors.
The metric can be accuracy, recall, precision, or F1-score for classification
tasks.\vspace{-0.1cm}

\subsection{Blockchain Model\vspace{-0.1cm}}

We consider a permissionless blockchain where any individual can participate
in the maintenance of the blockchain network, and on-chain data are
accessible to all participants. The blockchain network is an open
peer-to-peer network of numerous miners, who execute ML tasks and
propose new blocks to win rewards. We denote the blockchain network
with $N$ miners as ${\cal M}=\{M_{1},M_{2},...,M_{N}$\}, in which
each element is the identifier (ID) of a miner.

Blockchain $\mathsf{C}$ is  a chain of one-way linked blocks containing
payload data like transactions and smart contracts. A block is denoted
as $\mathsf{B}_{h}$, where $h$ means block height, the number of
blocks counted from the genesis block $\mathsf{B}_{0}$ to $\mathsf{B}_{h}$.
The timestamp included in every block $\mathsf{B}_{i}$ is denoted
as $T(\mathsf{B}_{i})$. We represent $\mathsf{C}$ with a sequence
of blocks as: $\mathsf{C}=\{\mathsf{B}_{0}\rightarrow\mathsf{B}_{1}\rightarrow\cdots\rightarrow\mathsf{B}_{h}\}$,
where the subscript $h$ is the block height of $\mathsf{B}_{h}$,
i.e., the length of C. Each miner $M_{i}$ in the blockchain network
${\cal M}$ maintains its local replica of the blockchain, which is
the local chain of $M_{i}$ and denoted as $\mathsf{C}_{i}$. The
predefined rule that the miners comply with to append new blocks to
the chain is called a consensus protocol. Typically, a consensus
mechanism is composed of two major operations: block generation and
block validation. The latter operation takes a block as input and
is denoted as $\mathrm{Validate}(\cdot)$.

 In practice, network delay might incur race conditions where new
blocks fail to reach all the nodes in the network in time before another
block is generated. From a global perspective, the race conditions
cause asynchronous states in the blockchain network, resulting in
extra branches diverging from the main chain, i.e., forks \cite{Tschorsch2016}.
We denote the block on the main chain $\mathsf{C}$ at block height
$k$ as $\mathsf{B}_{k}$ and the fork at block height $k+1$ as $\widetilde{\mathsf{B}}_{k+1}$,
which links to the same block $\mathsf{B}_{k}$ as $\mathsf{B}_{k+1}$.
We assume every miner only has a local view of the blockchain and
extends the local chain according to the consensus protocol.

Cryptographic hash functions are extensively utilized in blockchain
systems to eliminate block tampering and Sybil attacks. They are usually
modeled as random oracles, which output a fixed-length value for every
unique input of arbitrary length, and are denoted as $\mathrm{Hash}(\cdot)$
in this paper. The input of $\mathrm{Hash}(\cdot)$ might be any data
object or the concatenation of multiple objects, symbolized by $\vert\vert$
between objects. Also, the hash value of the preceding block that
$\mathsf{B}$ points to is called the prehash of $\mathsf{B}$. \vspace{-0.1cm}

\subsection{Network Model\label{subsec:Communication-Model}\vspace{-0.1cm}}

We describe peer-to-peer networks with two operations: $\mathrm{Broadcast}_{{\cal M}}(\cdot)$
and $\mathrm{Fetch}(\cdot,\cdot)$. The operation $\mathrm{Broadcast}_{{\cal M}}(\cdot)$
propagates messages such as blocks and payload data across the entire
blockchain network ${\cal M}$ so that all the miners can receive
the message. The operation $\mathrm{Broadcast}_{{\cal M}}(\cdot)$
varies with different network models.  For example, in a fully-connected
network, when a node generates a new block $\mathsf{B}$ and diffuses
it to the rest of the network $\mathcal{M}$ via $\mathrm{Broadcast}_{{\cal M}}(\mathsf{B})$,
the other nodes in the network $\mathcal{M}$ will receive the newly
generated block $\mathsf{B}$ after a given constant delay. We also
consider the mesh network model, a harsher network model with sparse
connections, to simulate wireless scenarios. In the mesh network model,
the transmission delay of a data object between two nodes in the network
is directly proportional to the object size and inversely proportional
to the link bandwidth between the two nodes.

The operation $\mathrm{Fetch}(\cdot,\cdot)$ is for the point-to-point
transmission of model parameters or datasets. When node $M_{j}$ invokes
$\mathrm{Fetch}(M_{i},\mathrm{Hash}(\Theta))$, it sends a query to
node $M_{i}$ with the identifier $\mathrm{Hash}(\Theta)$ and downloads
the data object $\Theta$. We assume the mining network can support
downloading models and datasets within an acceptable delay through
$\mathrm{Fetch}(\cdot,\cdot)$.\vspace{-0.1cm}

\section{Framework of BagChain \label{sec:System-Design}}

\subsection{Blockchain Structure\vspace{-0.1cm}}

In this section, we show the proposed BagChain framework and how BagChain
realizes bagging-based distributed learning via a blockchain approach.
First, we design the blockchain structure of BagChain. To enable
requesters to publish ML tasks and disclose validation and test datasets,
the data structure of ML tasks is defined as a seven-element tuple:
$\mathsf{TASK}=(\mathrm{Hash}(D_{T}),\mathrm{Hash}(D_{V}),$$\mathrm{Hash}(D_{E}),\mathrm{Train}(\cdot;f),\mathrm{Aggregate}(\cdot),$\newline$\mathrm{Metric}(\cdot,\cdot),\mathrm{Metric}_{min})$.
The fourth to sixth items correspond to the script used for model
training, aggregation, and evaluation. The parameter $f$ in $\mathrm{Train}(\cdot;f)$
and the minimum requirement $\mathrm{Metric}_{min}$ for ML model
performance metrics are specified by the requester. Any requester
can publish $\mathsf{TASK}$ via a task publication transaction, which
further includes the model training fees, and the requester's digital
signature and ID. The ML task to be executed at block height $h$
is denoted as $\mathsf{TASK}_{h}$. A requester can publish the validation
or test dataset via the dataset publication message, which consists
of the current block height $h$, the type of the published dataset,
the timestamp of the publishing moment when the validation or test
dataset is published, the ongoing task ID $\mathrm{Hash}(\mathsf{TASK}_{h})$,
and the requester's digital signature.

Specifically, to integrate with the distributed bagging algorithm,
we design a three-layer blockchain structure consisting of MiniBlocks,
Ensemble Blocks, and Key Blocks. The structures and interconnections
of MiniBlocks, Ensemble Blocks, and Key Blocks are depicted in Fig.
\ref{fig:Overview-of-data-structure}.
\begin{itemize}
\item \emph{MiniBlock} ($\mathsf{MB}_{h}^{i}$) contains a unique base model
and records the ownership of the base model and the hash identifier
of its parameters $\mathrm{ModelHash}=\mathrm{Hash}(\boldsymbol{\omega}_{i}||M_{i})$.
\item \emph{Ensemble Block} ($\mathsf{EB}_{h}^{k}$) points to several MiniBlocks
to form an ensemble model and also includes the ensemble model's metric
$\mathrm{Metric}_{V}^{k}$ on the validation dataset $D_{V}$.
\item \emph{Key Block} ($\mathsf{KB}_{h}$) includes the ensemble model
rank,  task queue, and also blockchain payload data. In $\mathsf{KB}_{h}$,
$M_{\xi_{h}}$, $\mathrm{Metric}_{E}^{j}$, and $\mathrm{Metric}_{best}$
represent the ID of the producer of $\mathsf{KB}_{h}$, the metric
of $\mathsf{EB}_{h}^{j}$ on the test dataset $D_{E}$, and the metric
of the best ensemble model, respectively. \vspace{-0.15cm}
\end{itemize}

\subsection{Workflow\label{subsec:Workflow}\vspace{-0.1cm}}

\begin{figure}[t]
\begin{centering}
\includegraphics[scale=0.8]{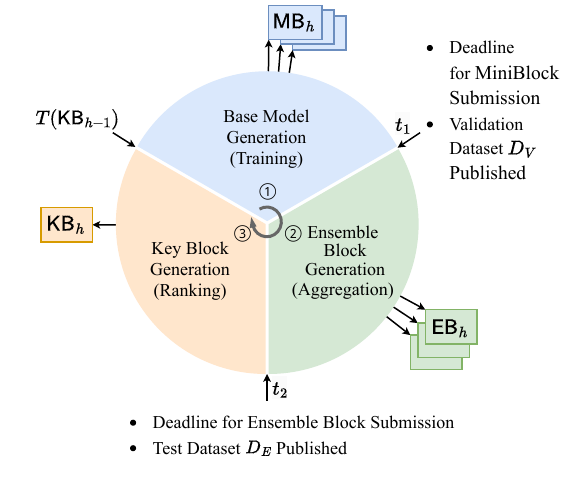}\vspace{-0.3cm}
\par\end{centering}
\caption{Workflow of BagChain and lifecycle of a learning task. \label{fig:Consensus-Process}}

\vspace{-0.4cm}
\end{figure}

The workflow of BagChain enables fully decentralized coordination
of miners and requesters. It includes  three phases, as illustrated
in Fig. \ref{fig:Consensus-Process}.  We assume the corresponding
task publication transaction has already been broadcast, validated
by miners, and recorded on the blockchain. 

\emph{Phase I: Base model training.} The generation and propagation
of a new Key Block $\mathsf{KB}_{h-1}$ trigger the first phase of
the execution of a new task, denoted as $\mathsf{TASK}_{h}$. If miner
$M_{i}$ receives multiple valid chains of different lengths, it will
extend the longest chain; otherwise, miner $M_{i}$ will extend the
chain with optimal $\mathrm{Metric}_{best}$.  Once miner $M_{i}$
confirms $\mathsf{KB}_{h-1}$, miner $M_{i}$ will execute Algorithm
\ref{alg:mb generation} to train a base model and generate a MiniBlock.
Miner $M_{i}$ first initiates the executable scripts and prepares
the local training dataset $D_{T_{i}}$, which includes its private
dataset $D_{M_{i}}$ and the public training dataset $D_{T}$. Then,
miner $M_{i}$ executes the training script on the bootstrapped local
dataset $D_{T_{i}}$ and generates the parameters of a base model,
denoted as $\boldsymbol{\omega}_{i}=\mathrm{Train}(\mathrm{Resample}(D_{T_{i}});f)$.
Once the base model is ready, miner $M_{i}$ encapsulates a timestamp
$T(\mathsf{MB}_{h}^{i})$, miner's ID $M_{i}$, $\mathrm{Hash}(\mathsf{TASK}_{h})$,
$\mathrm{ModelHash}$, and $\mathrm{Hash}(\mathsf{KB}_{h-1})$ into
$\mathsf{MB}_{h}^{i}$ and diffuses the MiniBlock via $\mathrm{Broadcast}_{{\cal M}}(\mathsf{MB}_{h}^{i})$.
All the miners should keep their base models secret  in Phase I to
avoid plagiarism.

\newcounter{subalg}
\setcounter{subalg}{0}
\renewcommand{\thealgorithm}{\arabic{algorithm}.\arabic{subalg}}

\begin{algorithm}[t]
\addtocounter{subalg}{1}\caption{MiniBlock Generation \label{alg:mb generation}}

\begin{algorithmic}[1]
\Require{$\mathsf{KB}_{h-1},M_i,D_{M_i}$}
\Ensure{$\mathsf{MB}_{h}^{i}$}

\State $\mathsf{TASK}_{h} \gets \mathsf{KB}_{h-1}$
\State $f,\mathrm{Train}(\cdot), \mathrm{Hash}(D_T) \gets \mathsf{TASK}_{h}$
\State $D_T \gets \mathrm{Fetch}(requester,\mathrm{Hash}(D_T))$
\State $D_{T_i} \gets D_T \bigcup D_{M_i}$
\State $\boldsymbol{\omega}_{i} \gets \mathrm{Train}(\mathrm{Resample}(D_{T_{i}});f)$
\State $\mathsf{MB}_{h}^{i} \gets \{ T(\mathsf{MB}_{h}^{i}) \vert \vert \mathrm{Hash}(\mathsf{TASK}_h) 
\vert \vert \mathrm{ModelHash} \vert \vert  M_i \vert \vert $
$\mathrm{Hash}(\mathsf{KB}_{h-1})\}$
\end{algorithmic}
\end{algorithm}

\emph{Phase II: Ensemble Block Generation.} Phase II starts at a pre-determined
time $t_{1}$. At $t_{1}$, the requester discloses the validation
dataset $D_{V}$ via a validation dataset publication message. Once
the validation dataset $D_{V}$ is released, any MiniBlock afterward
should be rejected since they may be trained by overfitting the validation
data. An honest miner $M_{i}$ will generate a new Ensemble Block
according to Algorithm \ref{alg:eb generation}, which can be summarized
into four steps. a) Download the validation dataset $D_{V}$ and execute
$\mathrm{Fetch}(M_{j},\mathrm{ModelHash})$ to fetch the parameters
$\boldsymbol{\omega}_{j}$, in which $\mathrm{ModelHash}$ is extracted
from $\mathsf{MB}_{h}^{j}$. b) Execute $\mathrm{Validate}(\mathsf{MB}_{h}^{j})$
for all the MiniBlocks that are collected by miner $M_{i}$ and point
to $\mathsf{KB}_{h-1}$ on the top of miner $M_{i}$'s local chain
$\mathsf{C}_{i}$. The MiniBlocks that are invalid or point to the
Key Blocks other than $\mathsf{KB}_{h-1}$ are discarded. c) Evaluate
the ensemble of all the available base models with the validation
dataset $D_{V}$ by calculating $\mathrm{Metric}_{V}^{i}=\mathrm{Metric}(\mathrm{Aggregate}($$f(\boldsymbol{X}_{V};\boldsymbol{\omega}_{1}),...,f(\boldsymbol{X}_{V};$$\boldsymbol{\omega}_{J})),\boldsymbol{y}_{V})$,
where $J$ is the total number of the valid MiniBlocks collected by
miner $M_{i}$. d) Encapsulate a timestamp $T(\mathsf{EB}_{h}^{i})$,
miner's ID $M_{i}$, $\mathrm{Hash}(\mathsf{TASK}_{h})$, $\mathrm{Metric}_{V}^{i}$,
and $\mathrm{Hash}(\mathsf{MB}_{h}^{j})$ into $\mathsf{EB}_{h}^{i}$.
The newly generated Ensemble Block $\mathsf{EB}_{h}^{i}$ is disseminated
by executing $\mathrm{Broadcast}_{{\cal M}}(\mathsf{EB}_{h}^{i})$.
 If miner $M_{i}$ receives any Ensemble Block $\mathsf{EB}_{h}^{q}$
from other miners, miner $M_{i}$ will verify $\mathsf{EB}_{h}^{q}$
with $\mathrm{Validate}(\mathsf{EB}_{h}^{q})$ (see Algorithm \ref{alg:eb validation})
and further forward $\mathsf{EB}_{h}^{q}$ to neighboring miners.

\begin{algorithm}[t]
\addtocounter{subalg}{1}
\addtocounter{algorithm}{-1}\caption{Ensemble Block Generation\label{alg:eb generation}}

\begin{algorithmic}[1]

\Require{$\mathsf{MB}_{h}^{1},...,\mathsf{MB}_{h}^{\hat{J}},M_i,\mathsf{KB}_{h-1}$}
\Ensure{$\mathsf{EB}_{h}^{i}$}

\State $\mathsf{TASK}_{h} \gets \mathsf{KB}_{h-1}$
\State $\mathrm{Aggregate}(\cdot) \gets \mathsf{TASK}_{h}$

\For {$j \gets 1$ to $\hat{J}$}
\If {$\mathrm{Validate}(\mathsf{MB}_{h}^{j})$ = False \textbf{or} prehash of $\mathsf{MB}_{h}^{j} \ne \mathrm{Hash}(\mathsf{KB}_{h-1})$} \label{CFS-Generate}
\State Discard $\mathsf{MB}_{h}^{j}$
\EndIf
\EndFor
\Comment $J$ out of $\hat{J}$ MiniBlocks are valid
\State $\mathrm{Metric}_{V}^{i} \gets \mathrm{Metric}(\mathrm{Aggregate}(f(\boldsymbol{X}_{V};\boldsymbol{\omega}_{1}),...,f(\boldsymbol{X}_{V};$
$\boldsymbol{\omega}_{J})),\boldsymbol{y}_{V})$
\State $ \mathsf{EB}_{h}^{i} \gets \{\mathrm{Hash}(\mathsf{MB}_{h}^{1}) \vert\vert \cdots \vert\vert \mathrm{Hash}(\mathsf{MB}_{h}^{J}) \vert\vert \mathrm{Metric}_{V}^{i} \vert\vert M_i \vert\vert $
$\mathrm{Hash}(\mathsf{TASK}_{h}) \vert\vert T(\mathsf{EB}_{h}^{i}) \}$

\end{algorithmic}
\end{algorithm}

\emph{Phase III: Key Block Generation.} Similar to Phase II, Phase
III starts at another pre-determined time $t_{2}$. At $t_{2}$, the
requester discloses the test dataset $D_{E}$. Once miner $M_{i}$
downloads the test dataset $D_{E}$, it verifies the Ensemble Blocks,
denoted by $\mathsf{EB}_{h}^{1},$$...,\mathsf{EB}_{h}^{K}$, and
evaluates them on the test dataset $D_{E}$, and attempts to create
a new Key Block via Algorithm \ref{alg:kb generation}. Then, miner
$M_{i}$ measures the performance of the Ensemble Blocks, e.g., $\mathsf{EB}_{h}^{j}$,
via $\mathrm{Metric}_{E}^{j}=\mathrm{Metric}($$\mathrm{Aggregate}(f(\boldsymbol{X}_{E};\boldsymbol{\omega}_{j,1}),...,f(\boldsymbol{X}_{E};\boldsymbol{\omega}_{j,\eta_{j}}))$$,\boldsymbol{y}_{E})$,
where $\boldsymbol{\omega}_{j,\ell}$ is the parameters of the base
model ensembled in $\mathsf{EB}_{h}^{j}$. Miner $M_{i}$ ranks all
the ensemble models from the Ensemble Blocks $\mathsf{EB}_{h}^{1},...,\mathsf{EB}_{h}^{K}$
and  finds the best model with the optimal $\mathrm{Metric}_{E}^{q}$
.  At last, miner $M_{i}$ deletes $\mathsf{TASK}_{h}$ from the
task queue and appends a new task to the tail of the task queue before
it puts the necessary components of a Key Block into a candidate block
$\widehat{\mathsf{KB}}_{h}$. Miner $M_{i}$ performs hash trials
to find a suitable nonce satisfying $\mathrm{Hash}(\widehat{\mathsf{KB}}_{h})<\text{Target}$
to generate a new Key Block $\widehat{\mathsf{KB}}_{h}$ and broadcast
it to the rest of the network via $\mathrm{Broadcast}_{{\cal M}}(\widehat{\mathsf{KB}}_{h})$.
In our paper, $\text{Target}$ is a threshold controlling the difficulty
of generating Key Blocks, similar to PoW.

\begin{algorithm}[t]
\addtocounter{subalg}{1}
\addtocounter{algorithm}{-1}\caption{Key Block Generation \label{alg:kb generation}}

\begin{algorithmic}[1]

\Require{$\mathsf{EB}_{h}^{1},...,\mathsf{EB}_{h}^{\hat{K}},M_i,\mathsf{KB}_{h-1}$}
\Ensure{${\mathsf{KB}}_h$}
\State $\mathsf{TASK}_{h} \gets \mathsf{KB}_{h-1}$
\State $\mathrm{Hash}(D_E) \gets \mathsf{TASK}_{h}$
\State $\boldsymbol{X}_{E},\boldsymbol{y}_{E} \gets D_{E} \gets \mathrm{Fetch}(requester,\mathrm{Hash}(D_E))$

\For{$j \gets 1$ \textbf{to} $\hat{K}$}
\State $\mathsf{MB}_{h}^{1},...,\mathsf{MB}_{h}^{\eta_{j}} \gets \mathsf{EB}_{h}^{j}$
\If {$\mathrm{Validate}(\mathsf{EB}_{h}^{j}) = $ False }
\State Discard $\mathsf{EB}_{h}^{j}$
\EndIf
\State $\mathrm{Metric}_{E}^{j} \gets \mathrm{Metric}(\mathrm{Aggregate}(f(\boldsymbol{X}_{E};\boldsymbol{\omega}_{j,1}),...,$
$f(\boldsymbol{X}_{E};\boldsymbol{\omega}_{j,\eta_{j}})),\boldsymbol{y}_{E})$
\EndFor
\Comment $K$ out of $\hat{K}$ Ensemble Blocks are valid

\State Nonce $\gets 0$
\State $\mathrm{Metric}_{best} \gets \mathrm{max}(\mathrm{Metric}_{E}^{1},...,\mathrm{Metric}_{E}^{K})$
\Repeat
\State ${\mathsf{KB}}_h \gets  \{\text{Nonce} \vert\vert \text{Merkle Root} \vert\vert T({\mathsf{KB}}_{h}) \vert\vert \mathrm{Metric}_{best} \vert\vert $
$\mathrm{Hash}(\mathsf{EB}_{h}^{1}) \vert\vert \mathrm{Metric}_{E}^{1} \vert\vert \cdots \vert\vert \mathrm{Hash}(\mathsf{EB}_{h}^{K}) \vert\vert \mathrm{Metric}_{E}^{K} \vert\vert M_i \vert\vert $
$\mathrm{Hash}(\mathsf{TASK}_{h}) \vert\vert \text{Task\ Queue} \vert\vert \mathrm{Hash}(\mathsf{KB}_{h-1})\}$
\Statex \Comment{$\mathrm{Metric}_{E}^{1} \ge \mathrm{Metric}_{E}^{2} \ge \cdots \ge \mathrm{Metric}_{E}^{K}$}
\State $\text{Nonce} \gets \text{Nonce} + 1$ 
\Until{$\mathrm{Hash}({\mathsf{KB}}_h) < \text{Target}$}

\end{algorithmic}
\end{algorithm}

\vspace{-0.15cm}

\subsection{Incentive Mechanism\vspace{-0.1cm}}

 In BagChain, miners generating Key Blocks and MiniBlocks are rewarded
because the miners generate valid MiniBlocks and Key Blocks to prove
a certain amount of CPU or GPU resources has been consumed in ML model
training and evaluation.  Honest generators of Key Blocks conform
to the following rules to allocate rewards. 1) The training fees
from the requester are evenly distributed to all the miners producing
the MiniBlocks that the winning Ensemble Block points to. 2) The
Key Block reward is transferred to the generator of the Key Block.
Since the selection of the winning Key Block is partly dependent on
$\mathrm{Metric}_{best}$, every miner will hunt for the best ensemble
model by collecting as many Ensemble Blocks as possible in Phase III
to obtain the Key Block reward. Though the miner generating the winning
Ensemble Block will not be rewarded, the generators of MiniBlocks
will be eager to generate Ensemble Blocks in Phase II since they must
ensure the hash values of their MiniBlocks are included in at least
one of the Ensemble Blocks to earn the training fees from the requester.\vspace{-0.1cm}

\section{Block Validation \label{sec:Block-Validation}}

BagChain's openness and permissionless setting lower the access barrier
but lead to new challenges since the open environment is less trustworthy.
The block generation rules in Section \ref{subsec:Workflow} can hardly
be enforced on dishonest nodes or adversaries. For example, lazy nodes
can compromise learning performance by plagiarizing other nodes' models,
and malicious nodes can disrupt the distributed learning process.
Consequently, BagChain requires efficient validation mechanisms that
provide quality and integrity guarantees for the produced models.

In this section, we will show the validation procedures of MiniBlocks,
Ensemble Blocks, and Key Blocks, respectively. Before validating any
of the three blocks generated at block height $h$, we assume miner
$M_{i}$ has received $\mathsf{TASK}_{h}$, and extracted $f$, $\mathrm{Aggregate}(\cdot)$,
$\mathrm{Metric}(\cdot,\cdot)$, $\mathrm{Hash}(D_{V})$, and also
$\mathrm{Hash}(D_{E})$ from $\mathsf{TASK}_{h}$, and downloaded
the corresponding validation or test datasets. Every miner maintains
a local cache to store downloaded data objects. If the Ensemble Blocks,
MiniBlocks, or model parameters $\omega_{p}$ involved in the validation
procedures are not available in the local cache, miner $M_{i}$ will
execute $\mathrm{Fetch}(\cdot,\cdot)$ to download the necessary data
objects.\vspace{-0.1cm}

\subsection{MiniBlock\label{par:Validation-of-Miniblocks}\vspace{-0.1cm}}

After miner $M_{i}$ receives a new MiniBlock $\mathsf{MB}_{h}^{j}$,
it verifies $\mathsf{MB}_{h}^{j}$ with $\mathrm{Validate}(\mathsf{MB}_{h}^{j})$
shown in Algorithm \ref{alg:mb validation} through the following
steps:
\begin{enumerate}
\item Download the model parameters $\boldsymbol{\omega}_{j}$ and verify
whether $\mathrm{Hash}(\boldsymbol{\omega}_{j}\vert\vert M_{j})$
matches the hash value $\mathrm{ModelHash}$ in $\mathsf{MB}_{h}^{j}$.\label{enu:Download-the-model}
\item Check whether the corresponding $\mathrm{Metric}(f(\boldsymbol{X}_{V};\boldsymbol{\omega}_{j}),$\newline$\boldsymbol{y}_{V})$
surpasses  $\mathrm{Metric}_{min}$ of $\mathsf{TASK}_{h}$, where
$D_{V}=(\boldsymbol{X}_{V},\boldsymbol{y}_{V})$ is the validation
dataset specified in $\mathsf{TASK}_{h}$.\label{enu:Check performance}
\end{enumerate}
Our design aims to solidify the base models and their ownership in
the MiniBlocks immutably and also to guarantee that the performance
of the base models is acceptable. We meet requesters' performance
requirements with step \ref{enu:Check performance}, where any base
model with unacceptable performance on the newly released validation
dataset $D_{V}$ is invalidated. Also, we verify the integrity and
ownership of the base model with step \ref{enu:Download-the-model},
where the non-invertible hash function $\mathrm{Hash}(\cdot)$ can
detect any error in model parameters $\boldsymbol{\omega}_{j}$ or
manipulation of miner ID $M_{j}$. We also prevent lazy miners from
uploading other miners' base models to BagChain with a two-stage model
submitting scheme, where new MiniBlocks and corresponding base models
will be rejected by honest miners after the validation dataset is
disclosed. From this point on, the base models that can be aggregated
in Phase II are immutable\@.\vspace{-0.1cm}

\begin{algorithm}[tbh]
\setcounter{subalg}{1}\caption{MiniBlock Validation\label{alg:mb validation}}

\begin{algorithmic}[1]
\Require{$\mathsf{MB}_{h}^{j}$}
\Ensure{True \textbf{or} False}

\State $\mathrm{Hash}(\mathsf{KB}_{h-1}),M_j,\mathrm{ModelHash}, \mathsf{TASK}_{h} \gets \mathsf{MB}_{h}^{j}$
\State $f,\mathrm{Metric}(\cdot), \mathrm{Metric}_{min}, \mathrm{Hash}(D_V) \gets \mathsf{TASK}_{h}$
\State $D_{V} \gets \mathrm{Fetch}(requester,\mathrm{Hash}(D_V))$
\State $\boldsymbol{X}_{V},\boldsymbol{y}_{V} \gets D_{V}$

\State $\boldsymbol{\omega}_{j} \gets \mathrm{Fetch}(M_{j},\mathrm{ModelHash})$
\If {$\mathrm{Hash}(\boldsymbol{\omega}_{j}\vert\vert M_{j}) \neq \label{ownership validation} \mathrm{ModelHash}$}
\State \Return False
\EndIf
\If {$\mathrm{Metric}(f(\boldsymbol{X}_{V};\boldsymbol{\omega}_{j}),\boldsymbol{y}_{V}) > \mathrm{Metric}_{min}$ 
}
\State \Return True
\Else
\State \Return False
\EndIf

\end{algorithmic}
\end{algorithm}

\subsection{Ensemble Block\label{par:Validation-of-Ensemble}\vspace{-0.1cm}}

After miner $M_{i}$ receives a new Ensemble Block $\mathsf{EB}_{h}^{q}$,
it verifies $\mathsf{EB}_{h}^{q}$ with $\mathrm{Validate}(\mathsf{EB}_{h}^{q})$
shown in Algorithm \ref{alg:eb validation} through the following
steps:
\begin{enumerate}
\item Obtain all $\mathsf{MB}_{h}^{p}$ whose hash values are contained
in $\mathsf{EB}_{h}^{q}$ and execute $\mathrm{Validate}(\mathsf{MB}_{h}^{p})$
to ensure all the MiniBlocks are valid.\label{enu:Obtain-MBs}
\item Check whether $\mathsf{MB}_{h}^{p}$ points to $\mathsf{KB}_{h-1}$,
the Key Block at block height $h-1$ in the local chain $\mathsf{C}_{i}$.\label{enu:preblock restrict}
\item Check whether the asserted metric $\mathrm{Metric}_{V}^{q}$ surpasses
$\mathrm{Metric}_{min}$ of $\mathsf{TASK}_{h}$ and equals $\mathrm{Metric}(\mathrm{Aggregate}($\newline$f(\boldsymbol{X}_{V};\boldsymbol{\omega}_{1}),...,f(\boldsymbol{X}_{V};$$\boldsymbol{\omega}_{G_{h,q}})),\boldsymbol{y}_{V})$
, where $G_{h,q}$ is the total number of MiniBlocks that $\mathsf{EB}_{h}^{q}$
points to.\label{enu:ensemble-model-evaluation}
\end{enumerate}

Similar to the validation steps for MiniBlocks, the above steps are
designed to ensure the performance of the ensemble models, which is
a vital design goal of BagChain. However, BagChain's functionality
hinges on the security and trustworthiness of the base models recorded
on the blockchain. Therefore, prior to validation of the Ensemble
Block $\mathsf{EB}_{h}^{q}$, we safeguard the model aggregation process
and combat model plagiarism and other malicious attacks by first forcing
miners to download the base model parameters and verify the integrity,
ownership, and performance of the base models in step \ref{enu:Obtain-MBs}.
On the top of verified MiniBlocks, we must guarantee the ensemble
models are aggregated correctly with an acceptable accuracy. We first
check the miners that generate the Ensemble Block $\mathsf{EB}_{h}^{q}$
and all MiniBlocks $\mathsf{MB}_{h}^{p}$ are working on the same
task by forcing all MiniBlocks $\mathsf{MB}_{h}^{p}$ to extend along
the local chain $\mathsf{C}_{i}$ of miner $M_{i}$ at block height
$h-1$. To prevent the ensemble model from falling below the threshold
$\mathrm{Metric}_{min}$, we require miners to verify the performance
of the ensemble model on the validation dataset $D_{V}$.

\begin{algorithm}[t]
\addtocounter{subalg}{1}
\addtocounter{algorithm}{-1}\caption{Ensemble Block Validation \label{alg:eb validation}}

\begin{algorithmic}[1]

\Require{$\mathsf{EB}_{h}^{q}$}
\Ensure{True \textbf{or} False}

\State $\mathsf{MB}_{h}^{1},...,\mathsf{MB}_{h}^{G_{h,q}}, \mathrm{Metric}_V^q, \mathsf{TASK}_{h} \gets \mathsf{EB}_{h}^{q}$
\State $\mathrm{Aggregate}(\cdot) \gets \mathsf{TASK}_{h}$

\For {$p \gets 1$ to $G_{h,q}$}
\If {$\mathrm{Validate}(\mathsf{MB}_{h}^{p})=$ False \textbf{or} prehash of $\mathsf{MB}_{h}^{p} \ne \mathrm{Hash}(\mathsf{KB}_{h-1})$} \label{CFS-Validate}
\State \Return False
\EndIf
\EndFor

\If {$\mathrm{Metric}_V^q = \mathrm{Metric}(\mathrm{Aggregate}(f(\boldsymbol{X}_{V};\boldsymbol{\omega}_{1}),...,f(\boldsymbol{X}_{V};$
$\boldsymbol{\omega}_{G_{h,q}})),\boldsymbol{y}_{V})$ \textbf{and} $\mathrm{Metric}_V^q > \mathrm{Metric}_{min}$}
\State \Return True
\Else
\State \Return False
\EndIf

\end{algorithmic}
\end{algorithm}
\vspace{-0.2cm}

\subsection{Key Block\label{par:Validation-of-Key-Blocks}\vspace{-0.1cm}}

After miner $M_{i}$ receives a new Key Block $\mathsf{KB}_{h}$,
it verifies $\mathsf{KB}_{h}$ with the operation $\mathrm{Validate}(\mathsf{KB}_{h})$
shown in Algorithm \ref{alg:kb validation} through the following
steps:
\begin{enumerate}
\item Retrieve $\mathsf{KB}_{h}$ and its preceding Key Block $\mathsf{KB}_{h-1}$
from the local chain $\mathsf{C}_{i}$. If $\mathsf{KB}_{h-1}$ is
included in the local chain $\mathsf{C}_{i}$, it should have been
verified and the following steps can be skipped. If $\mathsf{KB}_{h-1}$
is not included in $\mathsf{C}_{i}$, $\mathrm{Validate}(\mathsf{KB}_{h-1})$
should be executed first. 
\item Check whether $\mathrm{Hash}(\mathsf{KB}_{h})<\text{Target}$ holds.
\item Retrieve the optimal Ensemble Block $\mathsf{EB}_{h}^{q}$  according
to the hash pointer corresponding to $\mathrm{Metric}_{best}$ in
the metric list of $\mathsf{KB}_{h}$.
\item Execute $\mathrm{Validate}(\mathsf{MB}_{h}^{p})$ to verify all the
MiniBlocks in $\mathsf{EB}_{h}^{q}$ and whether they point to $\mathsf{KB}_{h-1}$.\label{enu:KB validate 1}
\item Aggregate the base models and evaluate the ensemble model with the
test dataset $D_{E}$ by calculating $\mathrm{Metric}=\mathrm{Metric}(\mathrm{Aggregate}(f(\boldsymbol{X}_{E};\boldsymbol{\omega}_{1}),...,f(\boldsymbol{X}_{E};\boldsymbol{\omega}_{G_{h,q}})),$$\boldsymbol{y}_{E})$.
\item Verify whether $\mathrm{Metric}=\mathrm{Metric}_{E}^{q}=\mathrm{Metric}_{best}$,
where $\mathrm{Metric}_{E}^{q}$ is the metric of $\mathsf{EB}_{h}^{q}$
in the metric list of $\mathsf{KB}_{h}$.
\item Repeat step 4 and 5 for all non-optimal Ensemble Blocks $\{\mathsf{EB}_{h}^{r}\}_{r=1}^{F_{h}}$
in the metric list of $\mathsf{KB}_{h}$ and verify whether $\mathrm{Metric}=\mathrm{Metric}_{E}^{r}\le\mathrm{Metric}_{best}$.\label{enu:KB validate 2}
\item Verify whether the payload data are valid and the rewards are correctly
allocated by traversing the Merkle Tree with root hash $\mathrm{Merkle}\ \mathrm{Root}$.\label{enu:KB validate 3}
\end{enumerate}
\begin{algorithm}[t]
\addtocounter{subalg}{1}
\addtocounter{algorithm}{-1}\caption{Key Block Validation\label{alg:kb validation}}

\begin{algorithmic}[1]
\Require{$\mathsf{KB}_{h}$}
\Ensure{True \textbf{or} False}

\State $\mathsf{EB}_{h}^{1},...,\mathsf{EB}_{h}^{F_{h}}, \mathrm{Metric}_{best}, \mathsf{TASK}_{h} \gets \mathsf{KB}_{h}$
\State $\mathrm{Aggregate}(\cdot), \mathrm{Hash}(D_E) \gets \mathsf{TASK}_{h}$
\State $D_{E} \gets \mathrm{Fetch}(requester,\mathrm{Hash}(D_E))$
\State $\boldsymbol{X}_{E},\boldsymbol{y}_{E} \gets D_{E}$

\If {
$\mathrm{Hash}(\mathsf{KB}_{h}) \ge \text{Target}$}
\State \Return False
\EndIf

\For {$r \gets 1$ to $F_{h}$}
\State $\mathsf{MB}_{h}^{1},...,\mathsf{MB}_{h}^{G_{h,r}} \gets \mathsf{EB}_{h}^{r}$

\For {$p \gets 1$ to $G_{h,r}$}
\If {$\mathrm{Validate}(\mathsf{MB}_{h}^{p})=$ False \textbf{or} prehash of $\mathsf{MB}_{h}^{p} \ne \mathrm{Hash}(\mathsf{KB}_{h-1})$}
\State \Return False
\EndIf
\EndFor

\If {$\mathrm{Metric}_E^r \ne \mathrm{Metric}(\mathrm{Aggregate}(f(\boldsymbol{X}_{E};\boldsymbol{\omega}_{1}),...,$
$f(\boldsymbol{X}_{E};\boldsymbol{\omega}_{G_{h,r}})),\boldsymbol{y}_{E})$ \textbf{or} $\mathrm{Metric}_E^r > \mathrm{Metric}_{best}$}
\State \Return False
\EndIf

\EndFor
\If {$\mathrm{Metric}_{best}$ in $\left\{ \mathrm{Metric}_E^r \right\}_{r=1}^{F_{h}}$ \textbf{and} payload data in the Merkle Tree are valid}
\State \Return True
\Else
\State \Return False
\EndIf
\end{algorithmic}
\end{algorithm}

We build reliable and trustworthy data storage by designing rigorous
validation steps for Key Blocks, thereby enabling miners to upload
training outcomes and requesters to receive desired ML models even
in an untrustworthy environment. Basically, we ensure BagChain's consistency
and immutability with the PoW validation steps and the hash chain
structure of Key Blocks. Then, we distinguish BagChain from classic
PoW systems with the recursive validation of the Ensemble Blocks and
MiniBlocks. The recursive validation procedures ensure the correctness
and accuracy of all the ensemble models and their base models by traversing
BagChain's three-layer chain structure and verifying all the models
involved.

Moreover, we include the incentive mechanism in the validation procedures
of Key Blocks. Recall that the training rewards are evenly distributed
to the miners that generate the base models of the winning ensemble
model, we add steps that check training reward allocation and invalidate
Key Blocks that pick an incorrect winning ensemble block by verifying
the metric of the winning ensemble model against the claimed metric
$\mathrm{Metric}_{best}$.

All above, the validation procedures of Key Blocks, along with the
fork resolution rule that retains the Key Block with the optimal $\mathrm{Metric}_{best}$
in the local chain $\mathsf{C}_{i}$, enable requesters to download
the best possible ensemble model with the information recorded on
the main chain.\vspace{-0.1cm}

\section{Cross Fork Sharing\label{sec:Cross-Fork-Sharing}}

In practical networks, miners may have different blockchain visions
due to the propagation delay or malicious attacks, and they may generate
different Key Blocks and cause blockchain forking. Multiple Key Blocks
may lead miners to work on different forks and split the blockchain
network into disjoint partitions mining on separate Key Blocks. Since
miners strictly stick to the validation procedures in Section \ref{sec:Block-Validation},
the base models  are only available for model aggregation on the
top of the same Key Block and are discarded by the miners working
on another. Consequently, the miners in each partition can only aggregate
a limited portion of MiniBlocks, wasting base models and computing
power. It also degrades the performance of the final ensemble model,
as further shown in Fig. \ref{fig:Model-performance-MiniBlock-waste}
in the simulation. The above phenomenon is referred to as \textquotedbl computing
power splitting\textquotedbl{} in our paper. 

\begin{figure}[t]
\begin{centering}
\vspace{-0.0cm}
\par\end{centering}
\begin{centering}
\includegraphics[scale=0.57]{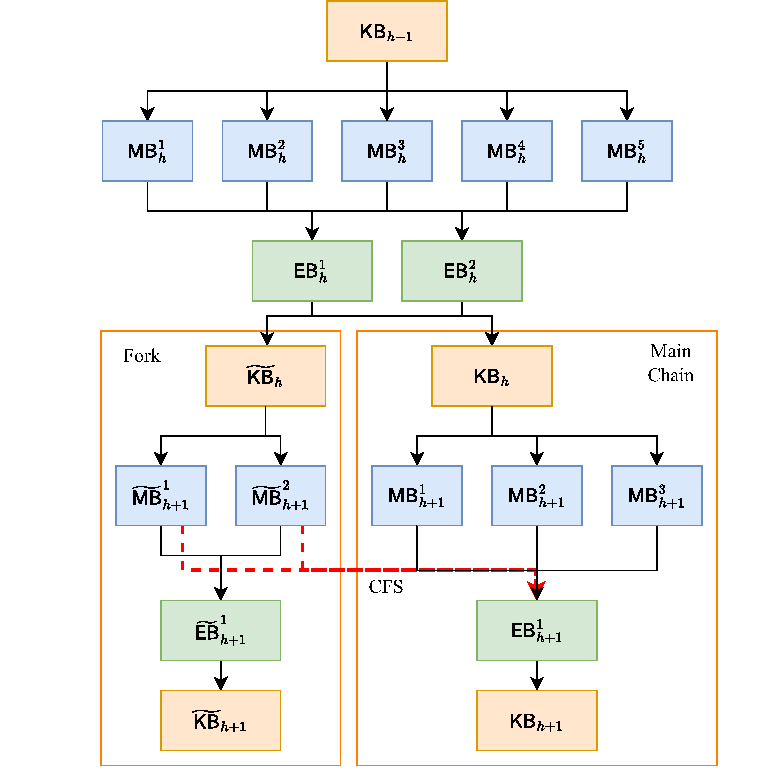}
\par\end{centering}
\begin{centering}
\vspace{-0.05cm}
\par\end{centering}
\centering{}\caption{Blockchain fork and the proposed cross fork sharing mechanism.\label{fig:Cross-Fork-Sharing}}
\vspace{-0.35cm}
\end{figure}

We visualize the above phenomenon in Fig. \ref{fig:Cross-Fork-Sharing}.
At block height $h$, there are two different Key Blocks, denoted
as $\widetilde{\mathsf{\mathsf{KB}}}_{h}$ and $\mathsf{KB}_{h}$,
and their child Key Blocks, denoted as $\widetilde{\mathsf{\mathsf{KB}}}_{h+1}$
and $\mathsf{KB}_{h+1}$. The Key Blocks $\mathsf{KB}_{h}$ and $\mathsf{KB}_{h+1}$
are on the main chain while $\widetilde{\mathsf{\mathsf{KB}}}_{h}$
and $\widetilde{\mathsf{\mathsf{KB}}}_{h+1}$ are forks. Note that
the Ensemble Block $\widetilde{\mathsf{\mathsf{EB}}}_{h+1}^{1}$ on
the fork points to two MiniBlocks and the Ensemble Block $\mathsf{EB}_{h+1}^{1}$
on the main chain points to three MiniBlocks. Hence, the ensemble
model corresponding to $\mathsf{EB}_{h+1}^{1}$ does not fully utilize
all the base models and thus the computing power splits at height
$h$. Obviously, the performance of the ensemble models is positively
related to the number of aggregated base models. As ``computing power
splitting'' occurs, some available computing power is wasted in generating
the base models on the fork and thus the performance of the final
ensemble model degrades. This is  why in Section \ref{subsec:Stale-Block-Rate}
the network delay affects the model accuracy on the test dataset.
Though the requester could download the base models from the miners
producing the two MiniBlocks $\widetilde{\mathsf{MB}}_{h+1}^{1}$
and $\widetilde{\mathsf{MB}}_{h+1}^{2}$ on the fork $\widetilde{\mathsf{\mathsf{KB}}}_{h+1}$
to improve ensemble model quality, these two MiniBlocks $\widetilde{\mathsf{MB}}_{h+1}^{1}$
and $\widetilde{\mathsf{MB}}_{h+1}^{2}$ are invalid on the main chain
$\mathsf{KB}_{h+1}$ according to Algorithm \ref{alg:kb validation}.
Also, the miners who generate $\widetilde{\mathsf{MB}}_{h+1}^{1}$
and $\widetilde{\mathsf{MB}}_{h+1}^{2}$ are unlikely to share their
base models with the requester if  they are not rewarded on the main
chain.

To make BagChain more robust to ``computing power splitting'', we
propose a cross fork sharing (CFS) mechanism to enable model sharing
among the main chain and different forks. In the CFS mechanism, we
remove the restrictions on the prehash of $\mathsf{\mathsf{MB}}_{h}^{p}$
in line \ref{CFS-Generate} of Algorithm \ref{alg:eb generation}
and line \ref{CFS-Validate} of Algorithm \ref{alg:eb validation}
and allow miners to leverage more base models they have received.
As depicted in Fig. \ref{fig:Cross-Fork-Sharing}, with the CFS mechanism,
the miner generating $\mathsf{EB}_{h+1}^{1}$ can aggregate the base
models on different forks such as $\widetilde{\mathsf{\mathsf{MB}}}_{h+1}^{1}$
and $\widetilde{\mathsf{\mathsf{MB}}}_{h+1}^{2}$ if the following
conditions are fulfilled:
\begin{enumerate}
\item All MiniBlocks to be included in $\mathsf{EB}_{h+1}^{1}$ should be
at the same block height and have the same task ID $\mathrm{Hash}(\mathsf{TASK}_{h+1})$.
Otherwise, the metrics of their base models can not be compared since
the test datasets of different tasks differ.\label{enu:CFS-condition2}
\item The aggregated base models should be unique in case any miner submits
duplicate base models on different forks.
\item The performance metrics of all aggregated base models must satisfy
the minimum requirement $\mathrm{Metric}_{min}$.
\end{enumerate}
Since different MiniBlocks might point to different Key Blocks at
the same block height, we employ a majority voting rule to determine
which Key block is on the main chain. Every valid MiniBlock is viewed
as a ``vote'' for the Key Block that the MiniBlock points to because
the MiniBlock represents a certain amount of time-consuming base model
training and Key Block validation work. Therefore, the hash value
of the parent Key Block should be derived from the majority voting
of the MiniBlocks that the winning Ensemble Block points to. For instance,
in Fig. \ref{fig:Cross-Fork-Sharing}, the prehash of $\mathsf{KB}_{h+1}$
should be $\mathrm{Hash}(\mathsf{KB}_{h})$ since among the MiniBlocks
that the winning Ensemble Block $\mathsf{EB}_{h+1}^{1}$ points to,
three MiniBlocks ``vote'' for $\mathsf{KB}_{h}$ while only two
MiniBlocks ``vote'' for $\widetilde{\mathsf{\mathsf{KB}}}_{h}$.
The performance of the CFS mechanism in high-latency networks is further
shown in Section \ref{subsec:Stale-Block-Rate}.\vspace{-0.1cm}

\section{Performance and Security\label{sec:Security-and-Perforance} }

\subsection{Underfitting and Overfitting\vspace{-0.1cm}}

In this section, we will dive deep into several performance and security
issues of BagChain. To prevent base models from underfitting, we introduce
the validation dataset and a threshold $\mathrm{Metric}_{min}$ in
BagChain. MiniBlocks will not be included in any valid Ensemble Block
unless the base models corresponding to the MiniBlocks satisfy the
metric threshold $\mathrm{Metric}_{min}$ of the ML task on the validation
dataset. In this way, unqualified base models are filtered out. 

Overfitting is a more important issue that needs to be particularly
considered for learning-based useful work proof. Lazy nodes may simply
train models by overfitting the samples in the test dataset, which
significantly compromises the learning performance. Therefore, in
BagChain, miners evaluate ensemble models on test datasets to minimize
the generalization error of the ensemble models of BagChain. In Phase
III (Key Block Generation), the requester discloses a test dataset
different from the validation dataset, and the miners evaluate all
the ensemble models collected in Phase II (Ensemble Block Generation)
on the test dataset. The miners overfitting the ensemble models on
the validation dataset reap no benefits since the ensemble models
overfitting on the validation dataset have no advantage over the other
ensemble models on the test dataset.\vspace{-0.2cm}

\subsection{Model Plagiarism\vspace{-0.1cm}}

\begin{figure}[t]
\begin{centering}
\vspace{-0.2cm}
\par\end{centering}
\begin{centering}
\includegraphics[scale=1.2]{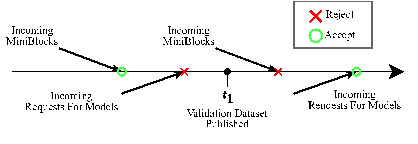}
\par\end{centering}
\begin{centering}
\vspace{-0.25cm}\caption{Model submitting scheme designed in BagChain workflow to combat ML
model plagiarism.\label{fig:Two-stage-model-committing}}
\par\end{centering}
\centering{}\vspace{-0.4cm}
\end{figure}

Model plagiarism is another important issue of PoUW protocols like
BagChain where lazy nodes may steal base models from other miners.
It discourages honest miners from contributing base models in BagChain.
 We address this issue via the model submitting scheme illustrated
in Fig. \ref{fig:Two-stage-model-committing}. The hash values of
the base models, i.e., $\mathrm{ModelHash}$, are submitted in MiniBlocks
first, and miners stop accepting MiniBlocks after the validation dataset
is published. The miners training the base models and generating MiniBlocks
do not accept incoming model download requests until the validation
dataset is published.  Hence, the lazy miners who attempt to steal
base models cannot download them from other miners before validation
dataset disclosure, and   any base model submitted with a MiniBlock
will be rejected by honest miners after the validation dataset is
published.

However, even with the model submitting scheme, lazy miners can copy
the hash value of $\boldsymbol{\omega}_{j}$ from miner $M_{j}$'s
MiniBlock as the $\mathrm{ModelHash}$ of their own MiniBlocks and
upload $\boldsymbol{\omega}_{j}$ as if $\boldsymbol{\omega}_{i}$
is their own base model after the validation dataset is published.
To close the loophole, we define $\mathrm{ModelHash}$ as $\mathrm{Hash}(\boldsymbol{\omega}_{i}||M_{i})$
to conceal the actual hash value of $\boldsymbol{\omega}_{i}$ and
bind the identity of miner $M_{i}$ to $\boldsymbol{\omega}_{i}$.
In this way, lazy miners' MiniBlocks containing a copied $\mathrm{ModelHash}$
and a mismatching miner ID will fail to pass the MiniBlock validation
step in line \ref{ownership validation} of Algorithm \ref{alg:mb validation}.\vspace{-0.2cm}

\subsection{Dataset Leakage Caused by Forking\label{subsec:Blockchain-Fork}\vspace{-0.1cm}}

\begin{figure}[t]
\begin{centering}
\subfloat[\label{fig:Dataset-Leakage1}]{\centering{}\includegraphics[scale=0.63]{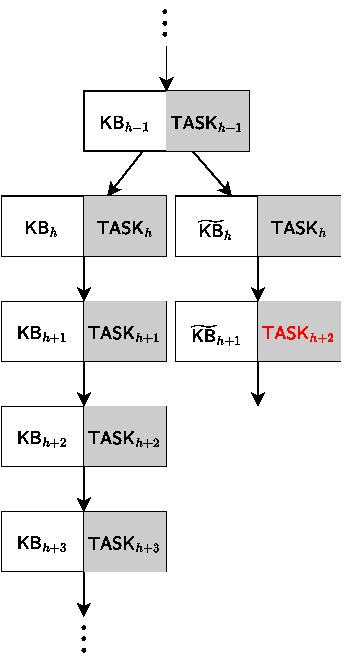}}\quad{}\subfloat[\label{fig:Dataset-Leakage2}]{\centering{}\includegraphics[scale=0.63]{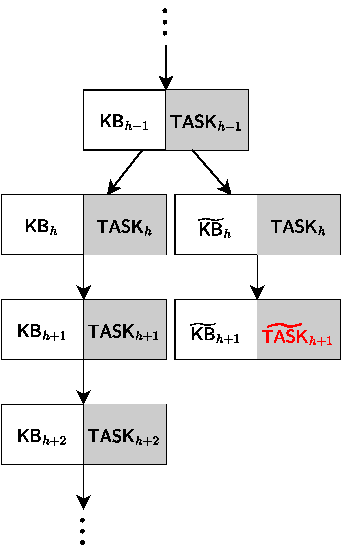}}\caption{Two possible situations when dataset leakage occurs in BagChain. (a)
Early execution of a task on the main chain. (b) Execution of a task
not recorded on the main chain. \label{fig:blockchain forks}}
\par\end{centering}
\centering{}\vspace{-0.4cm}
\end{figure}

As a distributed system, blockchain forking caused by network delays
or attackers may pose potential risks of leaking validation and test
datasets since two different subsets of miners might execute different
tasks at the same block height. Fig. \ref{fig:blockchain forks} depicts
the possible blockchain structures when blockchain forking occurs.
As shown in Fig. \ref{fig:Dataset-Leakage1} and \ref{fig:Dataset-Leakage2},
The two types of blockchain forks might lead to dataset leaks.

In Fig. \ref{fig:Dataset-Leakage1}, after $\mathsf{TASK}_{h}$ is
completed at block height $h$, a fork $\widetilde{\mathsf{KB}}_{h}$
is created. For the miners accepting $\widetilde{\mathsf{KB}}_{h}$
as the main chain, $\mathsf{TASK}_{h+2}$ is executed in advance,
resulting in early publication of the validation and test datasets
at block height $h+1$. Moreover, the miners accepting $\mathsf{KB}_{h}$
as the main chain may train base models for $\mathsf{TASK}_{h+2}$.
In this case, these miners may train the base models on the test dataset
published at block height $h+1$ for better scores on the test dataset.
However, they actually overfit the test dataset and are against the
intention of introducing validation and test datasets into BagChain.
If only a minority of miners train and aggregate models on the fork,
the produced ensemble model may underperform the ensemble model on
the main chain. Such deterioration in ensemble model performance on
the test dataset is undesired for requesters.

In Fig. \ref{fig:Dataset-Leakage2}, after $\mathsf{TASK}_{h}$ is
completed at block height $h$, $\widetilde{\mathsf{TASK}}_{h+1}$
is generated by the requester, included in the fork $\widetilde{\mathsf{KB}}_{h}$,
and executed at block height $h+1$. This means $\widetilde{\mathsf{TASK}}_{h+1}$
is executed before being confirmed in any Key Block on the main chain,
and thus its execution is not recorded on the main chain. As a result,
the training fees of $\widetilde{\mathsf{TASK}}_{h+1}$ are not deducted
from the requester's account on the main chain, and the miners working
on the fork $\widetilde{\mathsf{KB}}_{h}$ get nothing in return.

Consequently, concerned with the above situations, requesters will
hesitate to publish their datasets until they confirm that no fork
is created in the blockchain system, which delays dataset publication
and impairs BagChain's performance, especially in harsh network environments.

To tackle the issue, we set up a first-in-first-out queue in Key Blocks
to record upcoming tasks by storing their IDs. Whenever miners start
training base models, they pick up the first item in the task queue
as the current task. When generating Key Blocks, they remove the first
item from the original task queue and append a new ML task to the
tail of the new task queue. The new ML task can be either randomly
drawn from miners' task transaction pool or prioritized based on the
training fees offered by requesters. If a task enters a task queue
of length $Q$, the execution of the task will be scheduled $Q$ block
heights later, even if multiple forks are created. In this way, the
tasks executed on different forks at the same block height remain
unified. The choice of task queue length $Q$ is a trade-off since
$Q$ should be upper bounded to limit the size of Key Blocks and be
lower bounded to guarantee sufficient robustness to blockchain forking.

Last but not least, the task queue also benefits requesters in the
following aspects. First, with the task queue, they can predict the
tasks to be executed at the following block heights and prepare to
distribute the public training dataset in advance. Second, the task
queue can improve the effectiveness of the proposed CFS mechanism
and thus boost the performance of the ensemble models. Actually, as
long as the task queue is long enough, the task queue guarantees task
consistency across different blockchain forks and ensures the same
task is executed at the same block height, enabling cross-fork base
model sharing (see condition \ref{enu:CFS-condition2} in Section
\ref{sec:Cross-Fork-Sharing}). In this case, most base models generated
at the same block height can be aggregated into the Ensemble Block
in Phase II, which reduces computing power waste in BagChain.\vspace{-0.1cm}

\section{ChainXim-based Experiments\label{sec:ChainXim-based-Simulation}}

\subsection{Simulation Setup\vspace{-0.1cm}}

\begin{figure}[t]
\begin{centering}
\subfloat[\label{fig:Round-in-ChainXim}]{\begin{centering}
\includegraphics[scale=0.52]{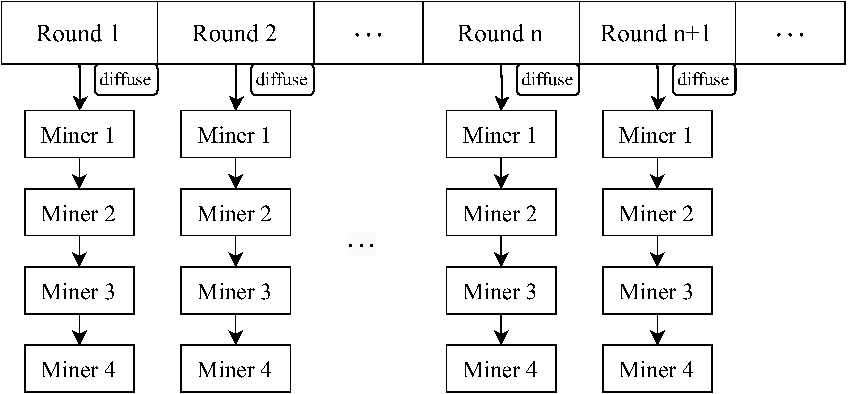}
\par\end{centering}
}
\par\end{centering}
\begin{centering}
\subfloat[\label{fig:topology-mesh-network}]{\begin{centering}
\includegraphics[scale=0.6]{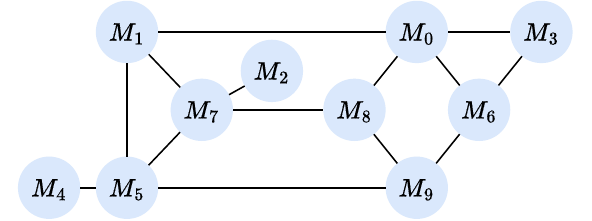}
\par\end{centering}
}
\par\end{centering}
\caption{Overview of ChainXim's internal mechanisms. (a) Discrete-time simulation
workflow utilizing \textquotedblleft round\textquotedblright{} as the
minimum time unit for simulating miners' actions and block dissemination.
(b) Topology of the mesh network used in Fig. \ref{fig:Model-performance-MiniBlock-waste}
and Fig. \ref{fig:Model-performance-upper-bound}.}
\vspace{-0.4cm}
\end{figure}

We realize and evaluate the framework of BagChain based on ChainXim,\footnote{ChainXim is available at https://github.com/ChainXim-Team/ChainXim.}
a blockchain simulator based on a discrete-time Byzantine setting
\cite{Garay2024}.  In ChainXim, time is segmented into many intervals
as ``\emph{rounds}''. As Fig. \ref{fig:Round-in-ChainXim} illustrates,
every miner validates the received blocks and merges valid ones into
the local chain. It attempts to generate new blocks to the main chain
in its own view and diffuses the newly generated blocks (if any) via
the network layer. ChainXim supports various user-configured network
models to simulate different scenarios. Specifically, the simulations
in Section \ref{subsec:Stale-Block-Rate} use the topology depicted
in Fig. \ref{fig:topology-mesh-network}. 

 Our simulations are based on several widely used image classification
tasks including MNIST \cite{Deng2012}, CIFAR-10 \cite{Krizhevsky2009},
SVHN \cite{Netzer2012}, and FEMNIST \cite{Caldas2019}. Our simulation
experiments not only consider various ML tasks but also different
types of base models including decision trees and a series of convolutional
neural networks (CNNs). Decision trees are classic ML classifiers
frequently used in conjunction with ensemble learning techniques like
bagging \cite{Breiman_1996} and random forest, and we use the implementation
in scikit-learn. The CNNs include LeNet \cite{Lecun1998}, ResNet18
\cite{He2016}, DenseNet \cite{Huang2017}, and MobileNetV2 \cite{Sandler2018}.
All neural networks are implemented with Pytorch and trained with
a learning rate of 0.001. 

For a given ML task, we construct the public training dataset $D_{T}$,
the validation dataset $D_{V}$, and the test dataset $D_{E}$ from
its original datasets.  We randomly select $\lvert\hat{D}_{E}\rvert/2$
samples in the original test dataset as the new validation dataset
$D_{V}$ in Phase II and set the rest as the new test dataset $D_{E}$
in Phase III. Furthermore, we split the original training dataset
$\text{\ensuremath{\hat{D}_{T}}}$ into a public training dataset
$D_{T}$ and several private datasets $D_{M_{i}}$ assigned to miners.
In the IID setting, we split the datasets evenly into homogeneous
subsets, whereas in the non-IID setting, we adopt the experimental
setup in \cite{Li2022may} to introduce distribution-based label heterogeneity
and real-world feature heterogeneity into the private datasets. More
details about dataset setups are presented in the Appendix.

The model accuracy, defined as the ratio of the correctly classified
samples to the total number of samples in a dataset, is used as the
performance metric to assess BagChain. During the simulations, ChainXim
records the accuracy of the ensemble models and the corresponding
base models on\textcolor{red}{{} }the test dataset $D_{E}$\textcolor{red}{{}
}at each block height. 

The size of MiniBlocks and Ensemble Blocks is set to 2 MB and the
size of Key Blocks is 6 MB by default. The corresponding propagation
delay can be calculated according to the bandwidth of the link. For
example, in the fully-connected network, it takes 12 rounds for a
block to reach all miners. In our simulation, Phase I and Phase II
are set to the fixed 100 rounds and 10 rounds, respectively. In other
words, the requester publishes the validation dataset 100 rounds after
Phase I starts and publishes the test dataset 10 rounds after Phase
II starts.

We set $\mathrm{Target=2^{244}-1}$ to safeguard the Key Block and
reduce the probability of forking. We implement $\mathrm{Hash}(\cdot)$
with the SHA256 hash function in ChainXim, so the probability that
a miner finds a valid nonce satisfying $\mathrm{Hash}(\widehat{\mathsf{KB}}_{h})<\text{Target}$
after $q$ mining attempts in each round is $1-(1-\mathrm{Target}/2^{256})^{q}$
($q=1$ in our simulation). Obviously, a smaller $\mathrm{Target}$,
i.e., a larger difficulty, reduces the forking probability but also
results in a higher Key Block mining delay.\vspace{-0.2cm}

\subsection{BagChain Performance for Different ML Tasks\label{subsec:Performance-Boost-BagChain}\vspace{-0.1cm}}

\begin{figure}[t]
\begin{centering}
\includegraphics[scale=0.45]{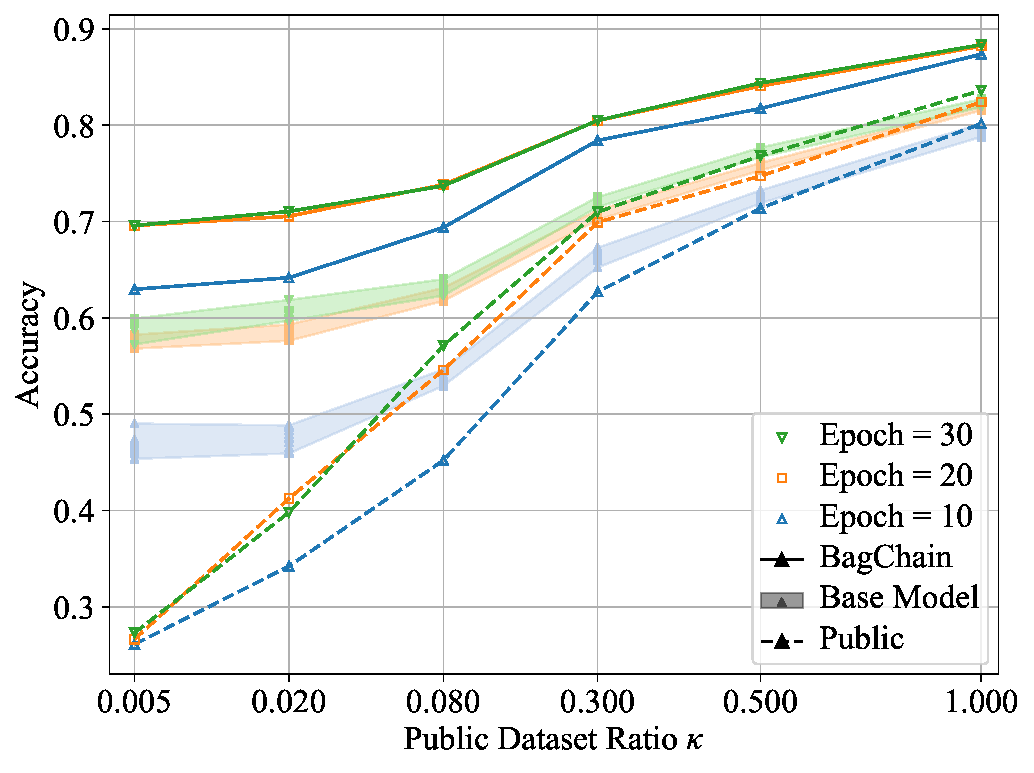}\vspace{-0.2cm}\caption{Performance boost of BagChain compared with base models in a fully-connected
network with $N=\varphi=10$. (The dataset is CIFAR-10 and the base
model is ResNet18.)\label{fig:base-ensemble-cifar10-iid}}
\par\end{centering}
\vspace{-0.4cm}
\end{figure}

In this section, we compare the performance of the ensemble models
generated by BagChain with that of the base models from the ChainXim-based
simulations in a fully connected network with 10 miners in an IID
setting.  As shown in Fig. \ref{fig:base-ensemble-cifar10-iid},
“BagChain”, ``Base Model'', and ``Public'' represent the ensemble
models based on the proposed BagChain, the base models trained on
every miner's local datasets $D_{T_{i}}$, and a dummy model directly
trained on the public training dataset $D_{T}$, respectively.  
The scattered points correspond to the accuracies of every miner's
local base models in the network, which is illustrated as the transparent
area to show the upper and lower bounds of local base models. The
different colors correspond to the cases where the base models are
trained for 10, 20, and 30 epochs, respectively.
\begin{table}[t]
\caption{Test dataset accuracy of the base and ensemble models generated in
BagChain\label{tab:validation-set-accuray}}

\begin{centering}
\par\end{centering}
\begin{centering}
\par\end{centering}
\begin{centering}
\begin{tabular}{|m{1.2cm}<{\centering}|m{1.7cm}<{\centering}|m{0.8cm}<{\centering}|m{1.8cm}<{\centering}|m{1.1cm}<{\centering}|}
\hline 
Dataset & Model & Public (\%) & Base Min\textasciitilde Max (\%) & BagChain (\%)\tabularnewline
\hline 
\hline 
MNIST & Decision Tree & 81.50 & 80.58\textasciitilde 81.92 & 91.66\tabularnewline
\hline 
\multirow{4}{*}{CIFAR-10} & LeNet & 50.48 & 44.48\textasciitilde 49.72 & 60.04\tabularnewline
\cline{2-5}
 & ResNet18 & 68.24 & 65.10\textasciitilde 70.22 & 76.70\tabularnewline
\cline{2-5}
 & DenseNet & 71.88 & 70.02\textasciitilde 73.76 & 81.58\tabularnewline
\cline{2-5}
 & MobileNetV2 & 60.70 & 58.46\textasciitilde 61.70 & 72.58\tabularnewline
\hline 
\multirow{4}{*}{SVHN} & LeNet & 80.65 & 81.75\textasciitilde 83.78 & 89.02\tabularnewline
\cline{2-5}
 & ResNet18 & 88.62 & 86.95\textasciitilde 92.03 & 96.00\tabularnewline
\cline{2-5}
 & DenseNet & 91.20 & 90.26\textasciitilde 92.11 & 95.29\tabularnewline
\cline{2-5}
 & MobileNetV2 & 89.38 & 86.21\textasciitilde 90.28 & 94.08\tabularnewline
\hline 
\multirow{4}{*}{FEMNIST} & LeNet & 77.02 & 74.70\textasciitilde 77.51 & 83.39\tabularnewline
\cline{2-5}
 & ResNet18 & 75.82 & 76.79\textasciitilde 80.31 & 85.07\tabularnewline
\cline{2-5}
 & DenseNet & 74.22 & 74.10\textasciitilde 81.10 & 85.82\tabularnewline
\cline{2-5}
 & MobileNetV2 & 76.13 & 75.63\textasciitilde 78.45 & 85.09\tabularnewline
\hline 
\end{tabular}
\par\end{centering}
\vspace{-0.2cm}
\end{table}
 
\begin{figure*}[t]
\begin{centering}
\vspace{-0.1cm}
\par\end{centering}
\begin{centering}
\subfloat[\label{fig:base-ensemble-mnist}]{\centering{}\includegraphics[scale=0.45]{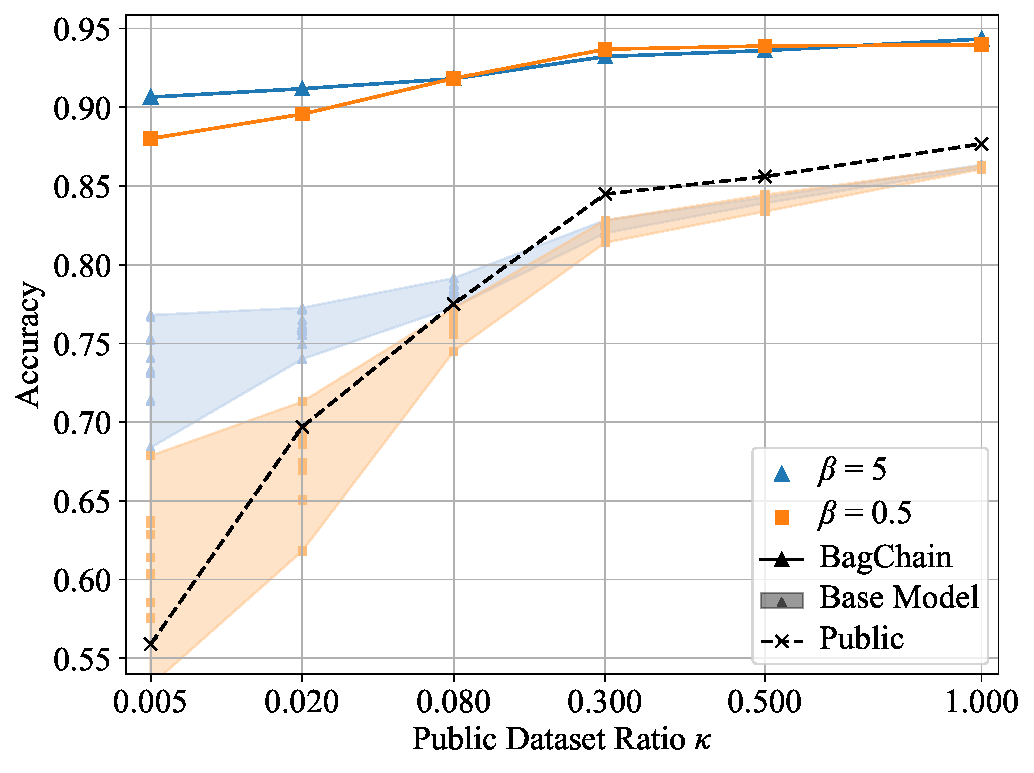}}\subfloat[\label{fig:base-ensemble-cifar10}]{\centering{}\includegraphics[scale=0.45]{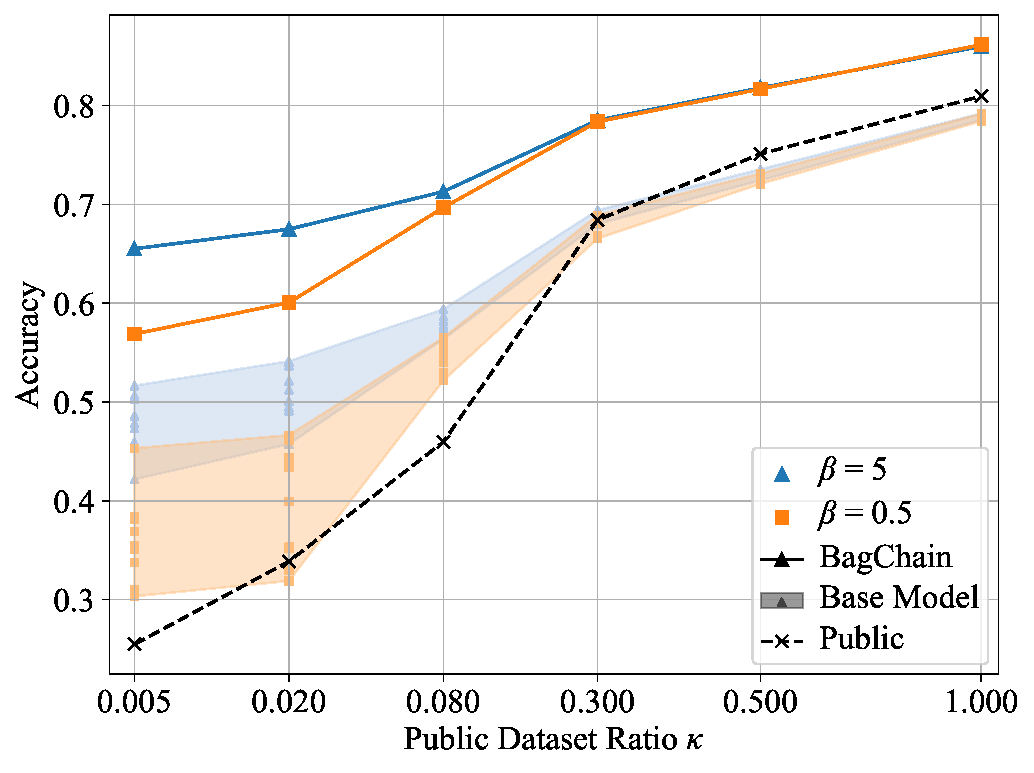}}
\par\end{centering}
\begin{centering}
\subfloat[\label{fig:base-ensemble-svhn}]{\centering{}\includegraphics[scale=0.45]{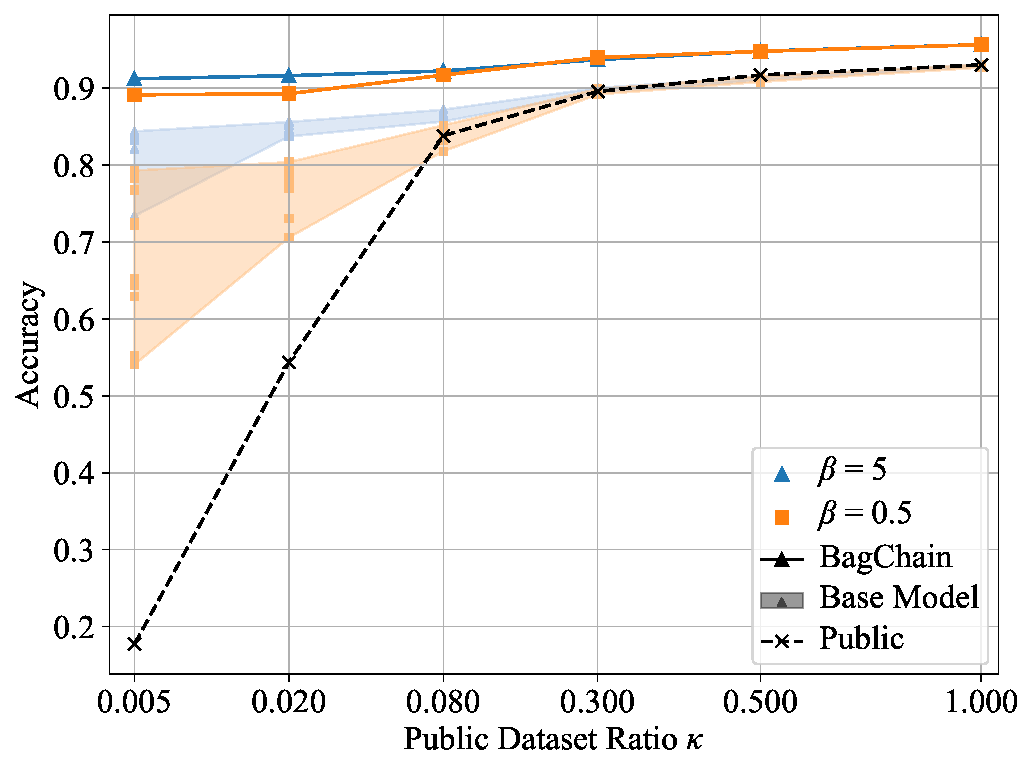}}\subfloat[\label{fig:base-ensemble-femnist}]{\centering{}\includegraphics[scale=0.45]{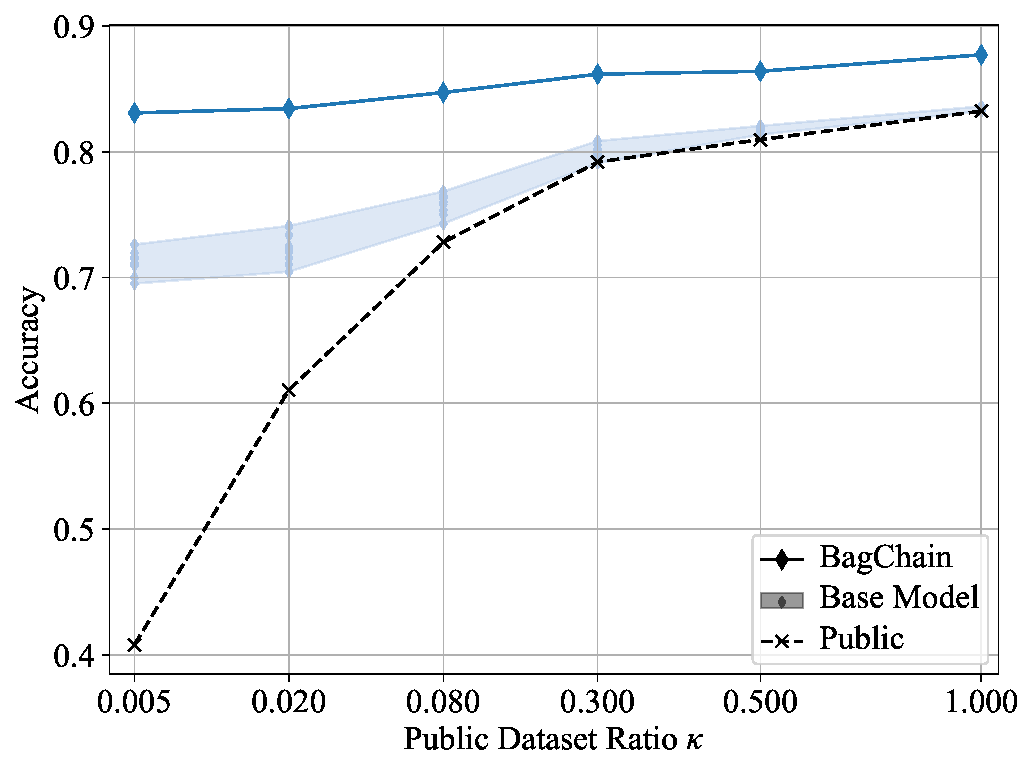}}
\par\end{centering}
\begin{centering}
\vspace{0.05cm}
\par\end{centering}
\caption{Performance boost of BagChain compared with base models under different
types and degrees of data heterogeneity in a fully-connected network
with N = $\varphi$ = 10. Every CNN is trained for 10 epochs. (a)
MNIST + Decision Tree. (b) CIFAR-10 + ResNet18. (c) SVHN + ResNet18.
(d) FEMNIST + ResNet18.\label{fig:base-ensemble}}

\centering{}\vspace{-0.5cm}
\end{figure*}

From Fig. \ref{fig:base-ensemble-cifar10-iid}, we observe that the
ensemble models outperform the base models trained on every miner's
local dataset and also the dummy model trained on the public dataset.
  This phenomenon highlights BagChain's capability to improve prediction
accuracy through the collective wisdom of multiple expert systems
trained on disjoint private and valuable datasets, thereby achieving
better performance than any single model trained on every miner's
private datasets or the requester's limited data samples.  Note that
BagChain does not require miners to reveal or exchange the raw dataset
for privacy protection. Not surprisingly, the accuracy of base and
ensemble models in BagChain increases as the public dataset ratio
$\kappa$ rises, i.e., more samples in the public training dataset
$D_{T}$. However, even when $\kappa=1$, i.e., all the models are
based on the same dataset, the ensemble model of BagChain still outperforms
other approaches. Therefore, apart from the knowledge from private
datasets, BagChain's accuracy-augmenting capability also benefits
from multiple miners' computing power and the performance gain of
the Bagging algorithm, i.e., aggregation of multiple expert models.
 Furthermore, we would like to highlight that BagChain does not rely
on well-trained base models. As shown in Fig. \ref{fig:base-ensemble-cifar10-iid},
around 20 epochs are sufficient to train base models since the ensemble
models corresponding to the base models trained by 20 and 30 epochs
exhibit a negligible performance gap. Hence, BagChain allows mobile
devices, which have local data samples but are resource-constrained,
to participate in model training without competing to fine-tune ML
models. Therefore, as shown in Fig. \ref{fig:base-ensemble-cifar10-iid},
BagChain can turn possibly weak base models into strong ensemble models
by utilizing distributed computing power and every node's local data
 without affecting individual privacy.

Furthermore, in Table \ref{tab:validation-set-accuray}, we illustrate
BagChain's capability and adaptability for various ML tasks using
different base models. In this experiment, BagChain is simulated in
a fully-connected network with the public dataset ratio $\kappa=0.1$.
 The column labeled \textquotedbl Base Min\textasciitilde Max (\%)\textquotedbl{}
provides the range of the accuracy of the miner's base models. From
Table \ref{tab:validation-set-accuray}, BagChain has about 3\%\textasciitilde 20\%
accuracy improvement compared with all the base models and the models
training solely with the public dataset for all tested models and
datasets. This demonstrates BagChain's generalization performance
in various ML tasks.\vspace{-0.2cm}

\subsection{BagChain Performance In the Non-IID Setting\vspace{-0.1cm}}

 In this section, we assess the performance of BagChain by taking
into account two different types of non-IIDness. In Fig. \ref{fig:base-ensemble-mnist}-\ref{fig:base-ensemble-svhn},
we construct label distribution heterogeneity in the datasets of MNIST,
CIFAR-10, and SVHN, respectively. In Fig. \ref{fig:base-ensemble-femnist},
we leverage the inherent feature heterogeneity in FEMNIST. The simulation
results consistently show the accuracy advantages of BagChain over
the base models and also the dummy \textquotedbl public\textquotedbl{}
model. Therefore, BagChain is effective and adaptable to different
non-IID ML tasks. Notably, Fig. \ref{fig:base-ensemble-femnist}
shows that BagChain performs effectively even when the base models
are trained on partitioned FEMNIST, where the features in each miner's
dataset are handwritten digits from distinct groups of writers, underscoring
BagChain's robustness to feature distribution heterogeneity.

Specifically, in Fig. \ref{fig:base-ensemble-mnist}-\ref{fig:base-ensemble-svhn},
we show the intensity of non-IIDness in private datasets across different
miners via the concentration parameter $\beta$. The concentration
parameter $\beta$ of the Dirichlet distribution $Dir(\beta)$ controls
the intensity of label distribution heterogeneity, and a smaller $\beta$
represents a higher degree of non-IIDness. The performance of base
models is significantly affected by the smaller concentration parameter
$\beta$, and meanwhile, the ensemble models generated by BagChain
remain robust against heterogeneity, highlighting BagChain's effectiveness
in the non-IID setting. Furthermore, from Fig. \ref{fig:base-ensemble-mnist}-\ref{fig:base-ensemble-svhn},
we observe that the gap between the two curves corresponding to $\beta=5$
and $\beta=0.5$ narrows as the public dataset ratio $\kappa$ increases.
This is because the public data samples help balance the label distribution
in miners' local datasets and reduce the error of each base model.

\begin{figure}[t]
\centering{}\includegraphics[scale=0.45]{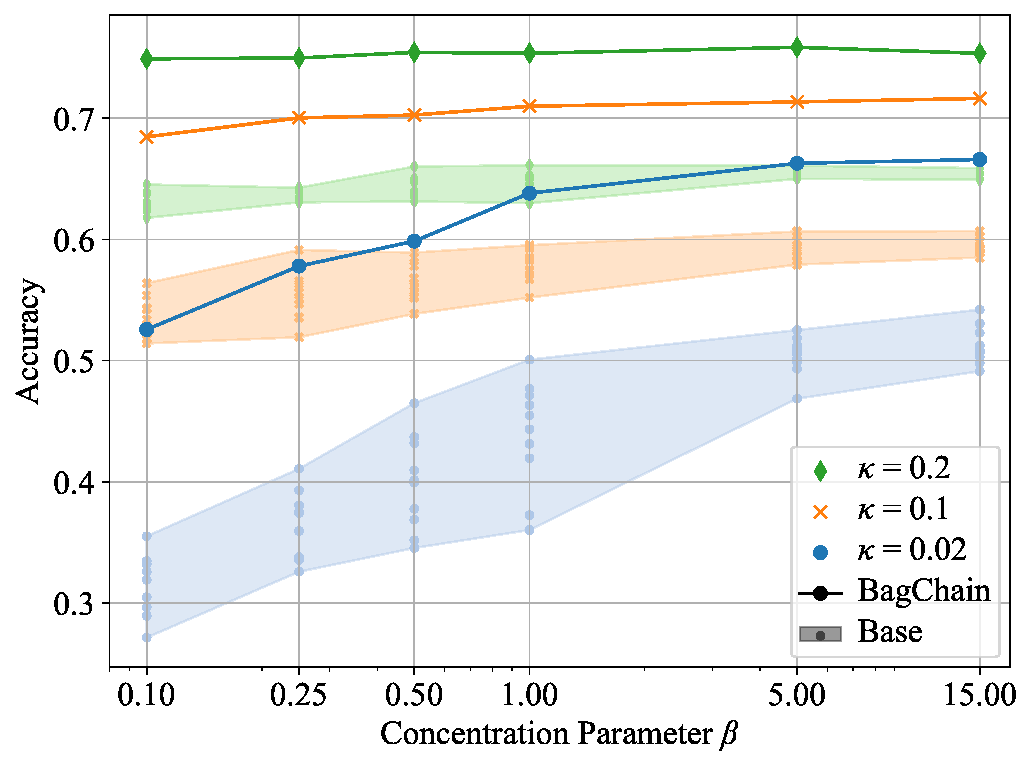}\vspace{-0.25cm}\caption{Model performance under different public dataset ratios $\text{\ensuremath{\kappa}}$
and concentration parameters $\beta$. (The dataset is CIFAR-10 and
the base model is ResNet18. Every base model is trained for 10 epochs.)
\label{fig:label-distr-cifar10}\vspace{-0.5cm}}
\end{figure}

We further illustrate BagChain's effectiveness under different degrees
of label distribution heterogeneity in Fig. \ref{fig:label-distr-cifar10}.
In this figure, the different colors correspond to different public
dataset ratios $\kappa=0.02,0.1,\text{and }0.2$, respectively. 
Similarly, all the ensemble models in BagChain achieve performance
advantage over the base models, even with severe label distribution
heterogeneity, indicating BagChain's effectiveness in the non-IID
setting. Also, as the public dataset ratio $\kappa$ increases from
0.02 to 0.2, the model accuracy curve flattens. When the public dataset
ratio $\kappa$ equals 0.2, we observe no significant accuracy drop
as the concentration parameter $\beta$ varies within the range {[}0.1,15{]}.
As explained above, this is because more public data helps reduce
the impact of label distribution heterogeneity and improve the performance
of both base models and ensemble models. Given the pervasive data
heterogeneity in real-world distributed learning scenarios with multiple
data sources, the above results demonstrate that BagChain is both
practical and effective for such applications.\vspace{-0.4cm}

\subsection{BagChain Robustness against Different Network Conditions\label{subsec:Stale-Block-Rate}\vspace{-0.1cm}}

\begin{figure}[t]
\centering{}\includegraphics[scale=0.45]{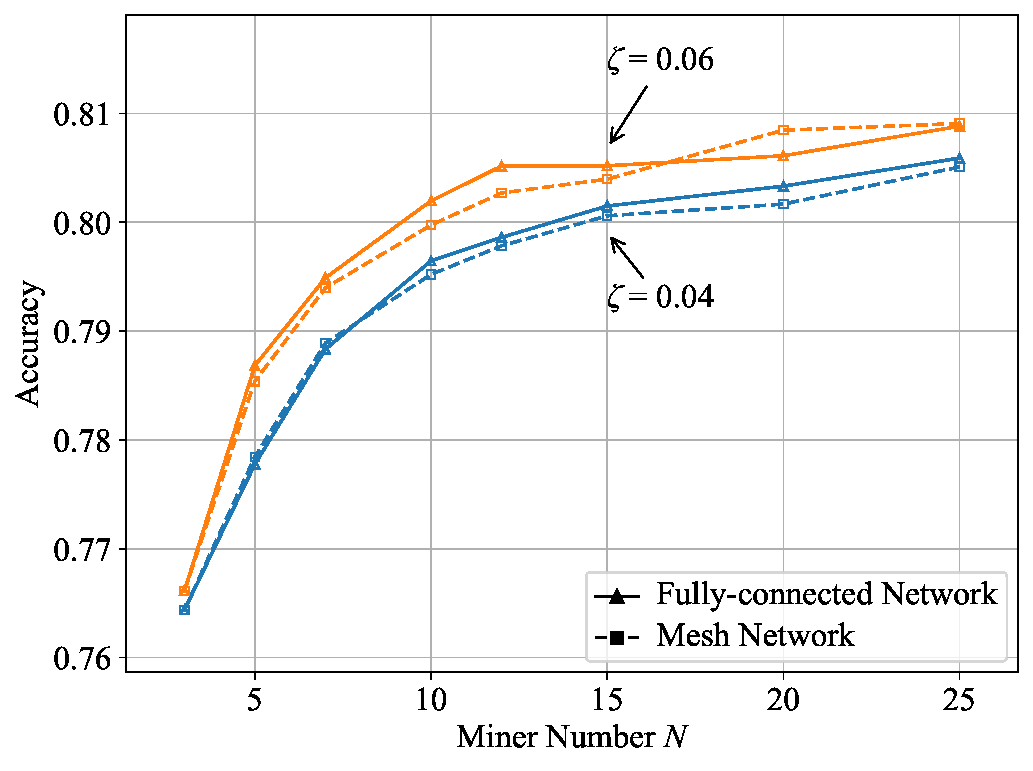}\vspace{-0.25cm}\caption{Model performance in different network scale. (The dataset is CIFAR-10
and the base model is ResNet18. Every base model is trained for 10
epochs and the public dataset ratio $\kappa$ equals 0.4.)\label{fig:Model-Performance-network-setting-cifar10}\vspace{-0.6cm}}
\end{figure}

In this section, we explore how network scale, network connectivity,
and the amount of private data impact the performance of BagChain.
In Fig. \ref{fig:Model-Performance-network-setting-cifar10}, the
x-axis ``Miner Number'' measures the scale of the blockchain network.
The ``Fully-connected Network'' and ``Mesh Network'' in the legend
are the fully-connected network and  mesh networks. In the fully-connected
network, each miner can communicate with all other miners. On the
contrary, in the mesh networks, each miner connects with only a subset
of the miners, and the topologies of the mesh networks are Erdős-Rényi
graphs randomly generated with networkx. In both network models, the
bandwidth between connected miners is 0.5 MB per round. The different
colors in Fig. \ref{fig:Model-Performance-network-setting-cifar10}
correspond to different private dataset ratios $\zeta=0.04\text{ and }0.06$,
which controls the number of samples in miners' private datasets.

Fig. \ref{fig:Model-Performance-network-setting-cifar10} shows that
BagChain is scalable since its accuracy increases with more miners.
This is because, as the network scales, more private data and computing
power are invested in training base models, and more base models are
aggregated. Also, a higher private dataset ratio $\zeta$ significantly
boosts the accuracy of the ensemble models generated in BagChain.
Hence, BagChain is more accurate if miners contribute more private
data to collaborative ML task training. Moreover, the negligible gap
between the accuracy in a fully connected network (solid lines) and
the one in a mesh network (dashed lines) in Fig. \ref{fig:Model-Performance-network-setting-cifar10}
indicates that BagChain experiences minor performance degradation
in a loosely connected blockchain network. Notably, even in a 25-node
mesh network with long-tailed propagation latency, the ensemble models
in BagChain remain accurate, highlighting its robustness against weak
connectivity and harsh network conditions.

\begin{figure}[t]
\begin{centering}
\vspace{-0.1cm}
\par\end{centering}
\centering{}\includegraphics[scale=0.45]{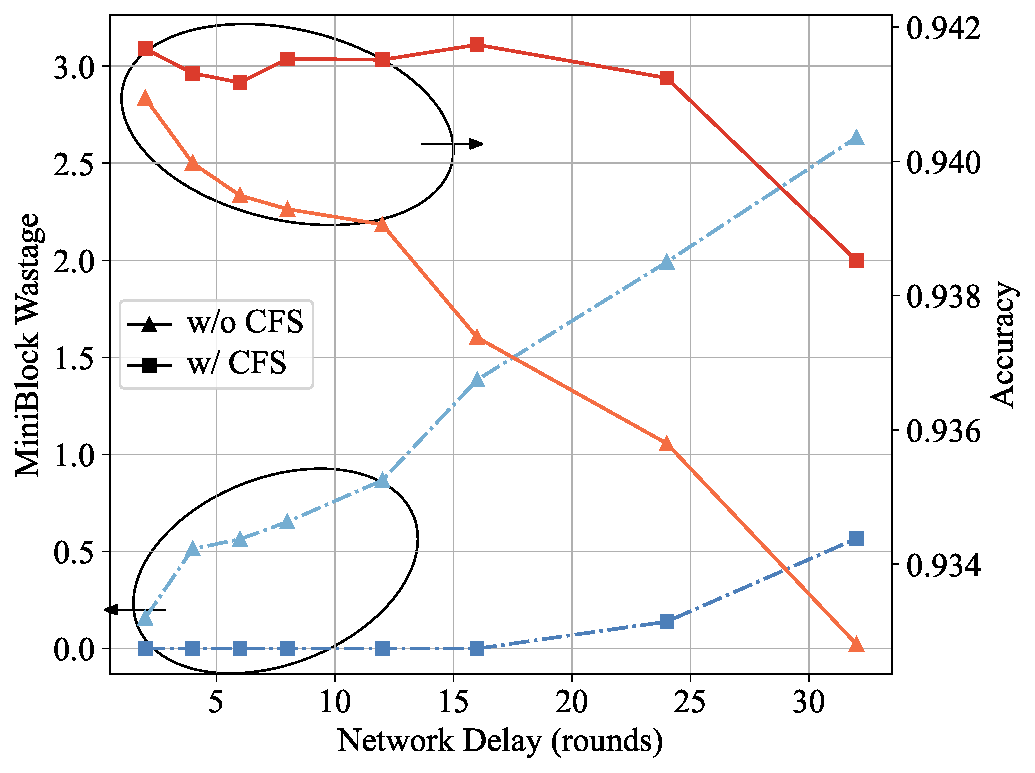}\vspace{-0.25cm}\caption{Average number of wasted MiniBlocks and accuracy of the ensemble models
generated in BagChain under different network delays. (The dataset
is MNIST and the base model is a decision tree.)\label{fig:Model-performance-MiniBlock-waste}\vspace{-0.0cm}}
\end{figure}

Furthermore, we investigate the performance of CFS by considering
blockchain forking to show the efficacy of the CFS mechanism. This
experiment uses a mesh network model with the topology depicted in
Fig. \ref{fig:topology-mesh-network}. As shown in Fig. \ref{fig:Model-performance-MiniBlock-waste},
“w/o CFS” represents the original BagChain scheme without CFS while
“w/ CFS” represents the improved BagChain scheme with CFS.  The x-axis
``Network Delay'' means the transmission delay for each Key Block
on each network connection. The y-axis ``MiniBlock Wastage'' is
 the average difference between the total number of MiniBlocks and
the number of MiniBlocks used by the winning Ensemble Block. 

Fig. \ref{fig:Model-performance-MiniBlock-waste} illustrates the
impact of network delay on MiniBlock wastage and the accuracy of the
ensemble models. From the curve marked ``w/o CFS'', we observe a
decrease in the accuracy of the ensemble models generated in ChainXim
as the network delay increases. Also, MiniBlock wastage rises and
ensemble model accuracy drops as the network latency increases. This
is because fewer base models can be utilized in model aggregation
in Phase II since frequent blockchain forking causes the ``computing
power splitting'' phenomenon described in Section \ref{sec:Cross-Fork-Sharing}.
In a blockchain network with high network delay, frequent blockchain
forking leads miners to generate MiniBlocks on different blockchain
forks, which cannot be shared among different forks. The crux of the
problem here is that the MiniBlocks a miner can utilize are limited
to those pointing to the same Key Block when generating an Ensemble
Block without using the CFS mechanism. It reduces the base models
available for model aggregation and causes computing power waste in
Phase II. In the curve marked ``w/ CFS'', the accuracy of the ensemble
models drops slightly below 0.94 as the network delay increases from
2 rounds to 32 rounds, and MiniBlock wastage is almost negligible
when the network delay is fewer than 32 rounds. This is because the
CFS mechanism improves BagChain by making all the MiniBlocks on different
forks available for model aggregation in Phase II. The more base
models are aggregated in ensemble models, the better performance the
ensemble models can achieve. Thereby, the proposed CFS mechanism can
improve ensemble model performance by avoiding computing power wastage
and maximizing the number of aggregated base models in ensemble models.

\begin{figure}[t]
\begin{centering}
\vspace{-0.1cm}
\par\end{centering}
\centering{}\includegraphics[scale=0.45]{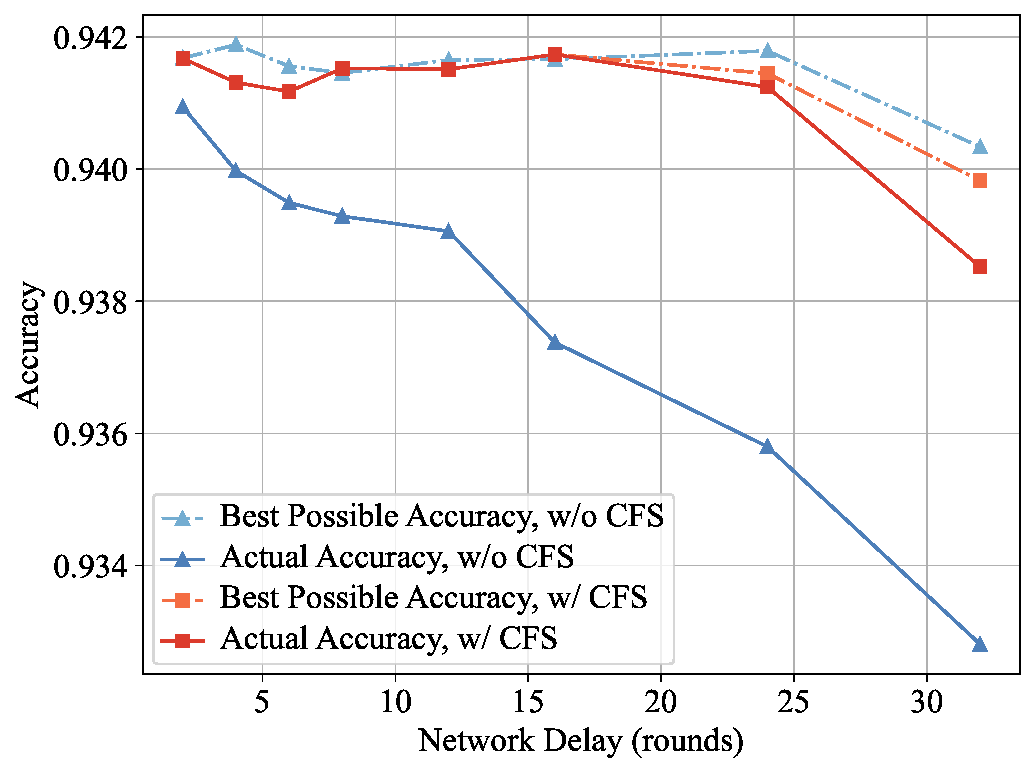}\vspace{-0.25cm}\caption{Actual accuracy and best possible accuracy of the ensemble models
generated in BagChain under different network delays. (The dataset
is MNIST and the base model is a decision tree.)\label{fig:Model-performance-upper-bound}\vspace{-0.4cm}}
\end{figure}

Fig. \ref{fig:Model-performance-upper-bound} further compares the
actual accuracy of the ensemble models with their best possible accuracy.
The actual accuracy is the accuracy of the winning ensemble models
in Key Blocks, while the best possible accuracy is the performance
metric of an ideal ensemble model aggregating all the base models
available for each task. The best possible accuracy is usually unattainable
in BagChain because the network delay may waste computing power (see
Section \ref{sec:Cross-Fork-Sharing}) or untimely exchange of MiniBlocks
and base models among miners. From Fig. \ref{fig:Model-performance-upper-bound},
compared with  BagChain without CFS, the actual accuracy of BagChain
with CFS  is closer to the best possible accuracy. The benefit of
CFS becomes more significant as the network delay increases.  Both
Fig. \ref{fig:Model-performance-MiniBlock-waste} and Fig. \ref{fig:Model-performance-upper-bound}
illustrate that the proposed CFS mechanism can utilize computing power
more efficiently and enhance BagChain's robustness to blockchain forking
and computing power splitting.\vspace{-0.1cm} 

\section{Conclusion\label{sec:Conclusion}}

In this paper, we propose and design BagChain, a bagging-based dual-functional
blockchain framework, to leverage both the computing power and private
data in a permissionless blockchain network. We record the results
of base model training, base model aggregation, and ensemble model
ranking in MiniBlocks, Ensemble Blocks, and Key Blocks, respectively,
and embed the bagging algorithm into the generation and verification
procedures of the corresponding three types of blocks.  We further
propose the CFS mechanism to maximize base model utilization, reduce
computing power waste, and improve BagChain's performance. We conduct
comprehensive experiments with ChainXim to demonstrate the efficacy
of BagChain for various ML tasks under different data heterogeneity
settings and network conditions. The simulation results in the IID
setting highlight the performance superiority of the ensemble models
in BagChain compared with the models trained solely with miners' restricted
datasets or requesters' public datasets. The experiments in the non-IID
setting demonstrate that the ensemble models generated in BagChain
can achieve desirable performance  even if the label and feature
distributions in miners' private datasets are heterogeneous. By testing
BagChain under different network conditions, we find that BagChain
is scalable and generates better ensemble models if larger computing
power and more private data samples are involved in base model training.
Thanks to the proposed CFS mechanism, BagChain exhibits strong robustness
against harsh network environments with poor connectivity and high
latency.\vspace{-0.1cm}

\appendix{}

\section{Dataset Setups\label{sec:Dataset-Creation}}

In the appendix, we would like to explain how we set up the datasets
in the simulations. The local dataset $D_{T_{i}}$ of miner $M_{i}$
is the union of the public training dataset $D_{T}$ and the private
dataset $D_{M_{i}}$. All miners share the same public training dataset
$D_{T}$ in each ML task, and the size of the public training dataset
$|D_{T}|$ is controlled by public dataset ratio $\kappa$, defined
as $\kappa=\lvert D_{T}\rvert/\lvert\hat{D}_{T}\rvert$. The way to
create the private dataset $D_{M_{i}}$ differs in IID and non-IID
settings.
\begin{table}[t]
\caption{Number of samples in each dataset split from the original training
dataset in the IID setting. (Miner number $N=\varphi=10$.) \label{tab:IID dataset}}

\centering{}%
\begin{tabular}{|m{4cm}<{\centering}|c|c|c|}
\hline 
Task & MNIST & CIFAR-10 & SVHN\tabularnewline
\hline 
\hline 
Total Training Samples $\vert\hat{D}_{T}\vert$ & 60000 & 50000 & 73257\tabularnewline
\hline 
Public Dataset Ratio $\kappa$ & \multicolumn{3}{c|}{0.4}\tabularnewline
\hline 
Public Dataset $\vert D_{T}\vert=\kappa\vert\hat{D}_{T}\vert$ & 24000 & 20000 & 29302\tabularnewline
\hline 
Private Dataset Ratio $\zeta=\vert D_{M_{i}}\vert/\vert\hat{D}_{T}\vert$ & \multicolumn{3}{c|}{0.06}\tabularnewline
\hline 
Private Dataset $\vert D_{P}\vert=\vert\hat{D}_{T}\vert-\vert D_{T}\vert$ & 36000 & 30000 & 43955\tabularnewline
\hline 
Private Dataset Per Miner $\vert D_{M_{i}}\vert=\vert D_{P}\vert/N$ & 3600 & 3000 & 4395\tabularnewline
\hline 
\end{tabular}
\end{table}

In the IID setting, we split the original training dataset $\hat{D}_{T}$
into a public part $D_{T}$ and a private part, which is then evenly
partitioned into $\varphi$ subsets, denoted as $D_{S_{i}}$ for $i=1,...,$$\varphi$.
(If $\varphi\ge N$, $D_{S_{i}}$ is assigned to miner $M_{i}$ ($D_{M_{i}}=D_{S_{i}}$)
for $i=1,...,N$. Otherwise, $D_{S_{j}}$ is assigned to miner $M_{i}$
($D_{M_{i}}=D_{S_{j}}$), in which $j$ equals $i$ for $i\le\varphi$
and is randomly selected from $\{1,2,...,\varphi\}$ for $i>\varphi$.)
Private dataset ratio $\zeta$ is defined as $\zeta=\lvert D_{S_{i}}\rvert/\lvert\hat{D}_{T}\rvert$,
controlling the size of miners' private datasets $|D_{M_{i}}|$. When
the public dataset ratio $\kappa$ is $0.4$ and private dataset ratio
$\zeta$ is 0.06, the number of samples in the datasets mentioned
above are shown in Table \ref{tab:IID dataset}.
\begin{figure}[t]
\begin{centering}
\vspace{-0.1cm}
\par\end{centering}
\begin{centering}
\includegraphics[scale=0.53]{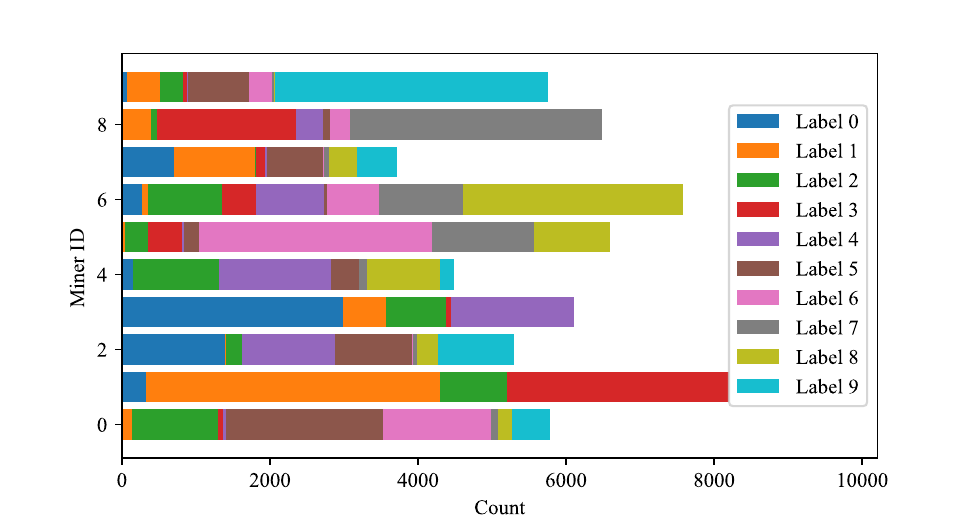}
\par\end{centering}
\begin{centering}
\vspace{-0.25cm}
\par\end{centering}
\centering{}\caption{Label distribution of each miner's private datasets when splitting
non-IID subsets from CIFAR-10 with $\beta=0.5$ and $N=10$.\label{fig:non-IID dataset}}
\vspace{-0.4cm}
\end{figure}

In the non-IID setting, we experiment with two distinct types of non-IIDness:
label distribution heterogeneity and feature distribution heterogeneity
\cite{Li2022may}. We curate datasets with heterogeneous label distributions
from balanced datasets including MNIST, CIFAR-10, and SVHN to simulate
label distribution heterogeneity, and we utilize the inherent heterogeneity
in the feature distribution of the data samples in FEMNIST to simulate
feature distribution heterogeneity. MNIST, CIFAR-10, and SVHN are
split into a public part $D_{T}$ and a private part $D_{P}$. Suppose
there are $L$ classes in the original training dataset $\hat{D}_{T}$,
and let the subset $D_{P_{k}}$ contain all samples of class $k$
in the private dataset $D_{P}$. We draw a probability mass function
$p_{k}:\{1,...,N\}\rightarrow[0,1]$ from the Dirichlet distribution
${Dir}_{N}(\beta)$ and then randomly partition the subset $D_{P_{k}}$
into $N$ subsets $\{D_{P_{k,j}}\}_{j=1}^{N}$, where each $D_{P_{k,j}}$
accounts for $p_{k}(j)$ of $D_{P_{k}}$ \cite{Zhu2021}. The private
datasets $D_{M_{j}}$ assigned to miner $M_{j}$ is the union of $\{D_{k,j}\}_{k=1}^{L}$.
The concentration parameter $\beta$ of the Dirichlet distribution
${Dir}_{N}(\beta)$ controls the degree of label distribution heterogeneity,
where a lower $\beta$ results in a higher degree of label imbalance.
As an example, in Fig. \ref{fig:non-IID dataset}, we depict the label
distribution of each miner's private dataset when splitting CIFAR-10
into non-IID subsets with concentration parameter $\beta=0.5$. For
FEMNIST, the training samples are the handwritten digits grouped into
non-overlapping subsets $D_{S_{i}}$ based on their writers. The
public dataset $D_{T}$ is constructed by randomly extracting $\lvert D_{T}\rvert/N$
samples from each $D_{S_{i}}$, and the private datasets $D_{M_{i}}$
contain the remaining samples in each subset $D_{S_{i}}$. \vspace{-0.1cm}

\bibliographystyle{IEEEtran}
\bibliography{202401survey,blockchain_learning_new,202309survey_new}

\end{document}